\documentclass[aps,rmp,reprint,amsmath,amssymb,graphicx,longbibliography]{revtex4-1}
\usepackage{bm}
\usepackage{color}
\usepackage{graphicx}
\usepackage{amsmath}
\usepackage{amssymb}
\usepackage{soul}
\usepackage[normalem]{ulem}

\makeatletter
\def\l@subsubsection#1#2{}
\makeatother

\newcommand{\ket}[1]{\ensuremath{\left| #1 \right\rangle}}
\newcommand{\bra}[1]{\ensuremath{\left\langle #1 \right|}}

\newcommand{\xiRG}{{\zeta_\text{RG}}}
\newcommand{\xiAB}{{\xi}}
\newcommand{\xiU}{{\xi}}
\newcommand{\xiOPE}{{\xi}}
\newcommand{\xiJ}{{\kappa}}
\newcommand{\xiE}{{\xi'}}

\newcommand{\ells}{{\ell_*}}

\def \be{\begin{equation}}
\def \ee{\end{equation}}

\usepackage{hyperref}
\hypersetup{
    bookmarks=false,         
    unicode=false,          
    pdftoolbar=false,        
    pdfmenubar=true,        
    pdffitwindow=false,     
    pdfstartview={FitH},    
    pdftitle={Many-body localization, thermalization, and entanglement},    
    pdfauthor={Dmitry A. Abanin, Ehud Altman, Immanuel Bloch, Maksym Serbyn },     
    pdfsubject={},   
    pdfcreator={},   
    pdfproducer={}, 
    pdfkeywords={many-body localization} {entanglement} {disordered systems}{quantum thermalization}, 
    pdfnewwindow=true,      
    colorlinks=true,       
    linkcolor=black,          
    citecolor=blue,        
    filecolor=magenta,      
    urlcolor=blue           
}

\begin{document}

\title{Many-body localization, thermalization, and entanglement}

\author{Dmitry A. Abanin}
\affiliation{Department of Theoretical Physics, University of Geneva, 1211 Geneva, Switzerland  }

\author{Ehud Altman}
\affiliation{Department of Physics, University of California, Berkeley, CA 94720, USA}

\author{Immanuel Bloch}
\affiliation{Fakult\"at f\"ur Physik, Ludwig-Maximilians-Universit\"at M\"unchen, 80799 Munich, Germany}
\affiliation{Max-Planck-Institut f\"ur Quantenoptik, 85748, Garching, Germany}

\author{Maksym Serbyn}
\affiliation{IST Austria, Am Campus 1, 3400 Klosterneuburg, Austria}

\date{\today}

\begin{abstract} 
Thermalizing quantum systems are conventionally described by statistical mechanics at equilibrium. However, not all systems fall into this category, with many body localization providing a generic mechanism for thermalization to fail in strongly disordered systems. Many-body localized (MBL) systems remain perfect insulators at non-zero temperature, which do not thermalize and therefore cannot be described using statistical mechanics. In this Colloquium we  review recent theoretical and experimental advances in studies of MBL systems,  focusing on the new perspective provided by entanglement and non-equilibrium experimental probes such as quantum quenches. Theoretically, MBL systems exhibit a new kind of robust integrability: an extensive set of quasi-local integrals of motion  emerges, which provides an intuitive explanation of the  breakdown of thermalization. A description based on quasi-local integrals of motion is used to predict dynamical properties of MBL systems, such as the spreading of quantum entanglement, the behavior of local observables, and the response to external dissipative processes. Furthermore, MBL systems can exhibit eigenstate transitions and quantum orders forbidden in thermodynamic equilibrium. We  outline the current theoretical understanding of the quantum-to-classical transition between many-body localized and ergodic phases, and anomalous transport in the vicinity of that transition. Experimentally, synthetic quantum systems, which are well-isolated from an external thermal reservoir, provide natural platforms for realizing the MBL phase. We review recent experiments with ultracold atoms, trapped ions, superconducting qubits, and quantum materials, in which different signatures of many-body localization have been observed. We conclude by listing outstanding challenges and promising future research directions. 
\end{abstract}

\maketitle

\tableofcontents{}

\section{Introduction \label{Sec:intro}}

Dramatic experimental progress of the last few years has enabled the realization of quantum many-body systems that are well isolated from the environment and therefore evolve under their intrinsic quantum dynamics. Examples of  systems that offer a large degree of control include ultracold atoms in optical lattices~\cite{BlochColdAtoms}, trapped ions~\cite{Blatt12} as well as nuclear and electron spins associated with impurity atoms in diamond~\cite{doherty2013nitrogen,Schirhagl2014}. The tunability and long coherence times of these systems, along with the ability to prepare highly non-equilibrium states, enable one to probe quantum dynamics and thermalization in closed systems. What are the possible regimes of quantum-coherent many-body dynamics? How does classical hydrodynamic transport, seen at long times when a system thermalizes, emerge from the unitary quantum evolution? Under what conditions does a system fail to thermalize, thus evading the conventional classical fate even at long times?  In contrast to a majority of experiments in solid state systems, these questions pertain to highly non-equilibrium states of matter with non-zero energy density that could translate to high and even infinite effective temperature. Can quantum effects survive at long times in many-body systems at such high energy densities? Answering these basic questions is a necessary step towards understanding a potentially very rich variety of new states of matter that can appear in highly non-equilibrium quantum systems. 

The most common class of dynamics leads to thermalization: in {\it ergodic}\footnote{We note that in the context of quantum many-body systems the term ergodicity is defined somewhat differently compared to classical mechanics. Our use of this term is synonymous with thermalization, as discussed in Section~\ref{Sec:MBL-Therm}.} systems, different degrees of freedom exchange energy and information. 
At long times, the system effectively reaches thermal equilibrium, even though as a whole it remains in a pure quantum state. Intuitively, in ergodic dynamics the system as a whole acts as a thermal reservoir for its subsystems, provided those are small enough. Stationary states in such systems are described by quantum statistical mechanics~\cite{DeutschETH,SrednickiETH}. 
 
\begin{figure}[b]
\begin{center}
\includegraphics[width=0.95\columnwidth]{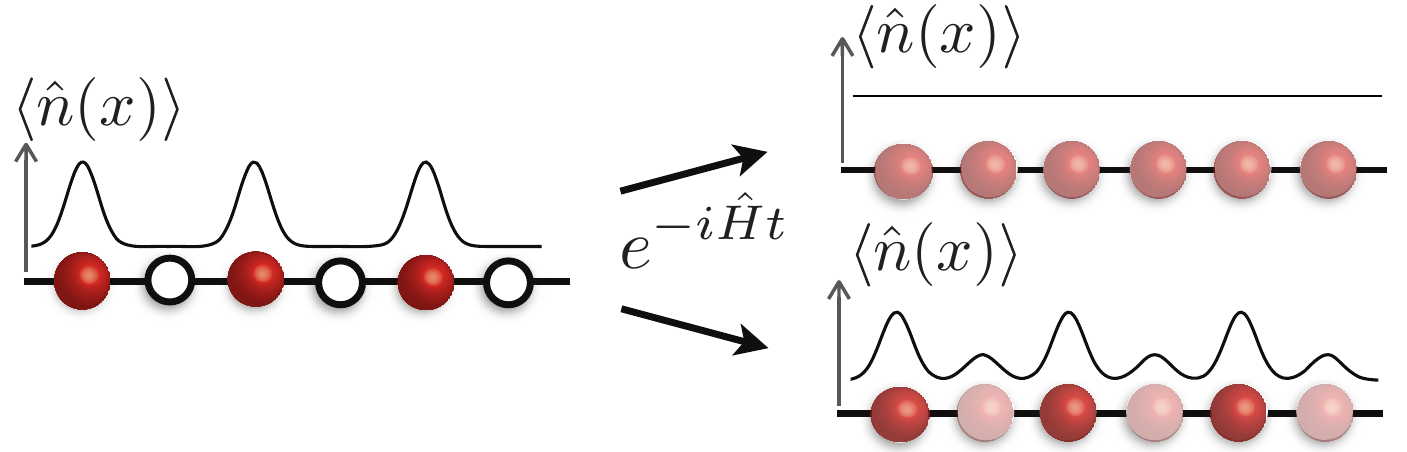}\\
\caption{ \label{Fig:1}
 In a quantum quench, interacting particles on a lattice are  e.g.\ initially prepared in a state with non-uniform density. Following unitary quantum dynamics, the thermalizing system relaxes towards the state where all lattice sites are equally populated and the density profile is uniform (shown at the top). In contrast, the many-body localized system retains the memory of initial state even at infinite time (bottom). }
\end{center}
\end{figure}

The approach to equilibrium is illustrated in Fig.~\ref{Fig:1} for a particular setup, known as a quantum quench, in which a system described by the Hamiltonian $\hat H$ is prepared in a non-equilibrium state $|\psi(0)\rangle$, e.g. characterized by a non-uniform density of particles. Under unitary evolution $e^{-i\hat{H}t}$, at sufficiently long times, the state $|\psi(t)\rangle=e^{-i\hat Ht}|\psi(0)\rangle$ of an ergodic system will have local observables which appear thermal. Information encoded in the initial state is effectively erased in the course of time evolution as it is transferred to highly non-local inaccessible  correlations. First to be washed away are the quantum correlations in the initial state, while the last to disappear are the inhomogeneities of conserved densities that are transported by slow diffusion modes. Ultimately the local physical observables will be determined  just by the values of the few global conserved quantities, total energy, particle number etc.

The mechanism of thermalization, as well the approach to thermal equilibrium in different systems, are issues of central importance in statistical mechanics. While there exist different regimes of thermalization (e.g., it can be parametrically slow), it is of particular interest to find systems which avoid thermalization.   In this case, quantum information encoded in the initial state can persist and govern the dynamics at long times as well as the steady state. Thus, ergodicity-breaking systems can allow for new forms of stable quantum phases and phase transitions that are unique to the non-equilibrium settings. Moreover, understanding ergodicity breaking mechanisms could  provide new insights into the workings of thermalization.

\begin{figure}[b]
\begin{center}
\includegraphics[width=0.8\columnwidth]{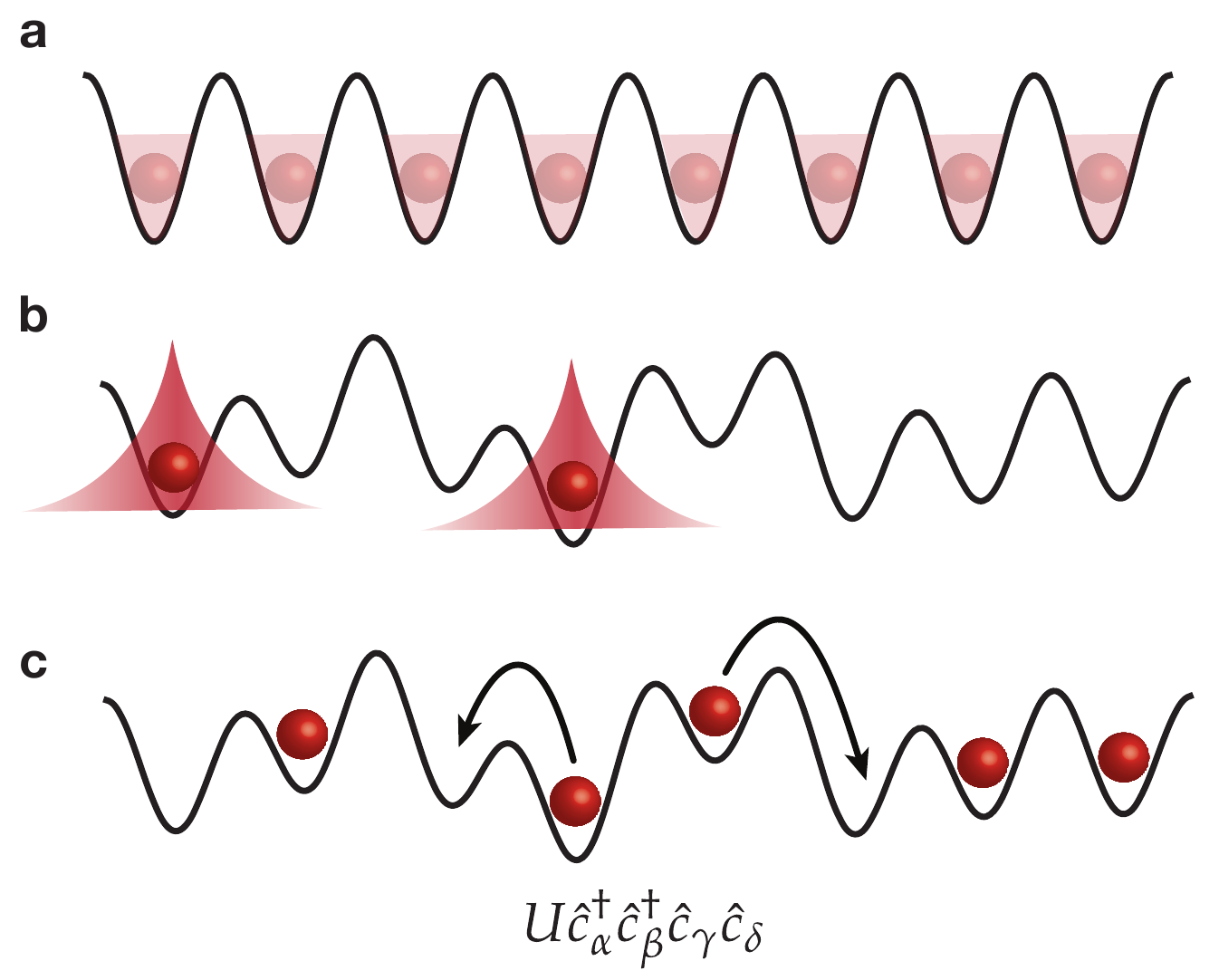}\\
\caption{ \label{Fig:Anderson}
(a) In a clean crystal, eigenstates are Bloch waves, which extend throughout the sample. 
(b) The essence of Anderson localization of non-interacting particles is that for sufficiently strong disorder there is a vanishing probability for a particle to make a resonant transition from one site to another one spatially separated from it. This leads to eigenstates which are localized in some region of space, decaying exponentially away from it. 
(c) Adding interactions to an Anderson localized system. To first order, the effect of interaction is to induce hopping of pairs of particles between the single particle localized orbitals. One may ask if the localized phase, with vanishing particle and thermal conductivities, is robust to this process. } 
\end{center}
\end{figure}

Thermalization requires that different parts of ergodic systems exchange energy and particles, and consequently thermal systems must be conducting. Therefore, a natural way to break ergodicity is to find systems which are insulating. One familiar and well-studied example of insulating behavior is Anderson localization in non-interacting disordered systems~\cite{Anderson58,50years}.  
The essence of Anderson localization is that a disorder potential can completely change the nature of single-particle eigenstates in a crystal: instead of propagating Bloch states, which are similar to plane waves (Fig.~\ref{Fig:Anderson}a), wave functions become localized in some region of space, and decay exponentially far away from that region (Fig.~\ref{Fig:Anderson}b). The origin of localization can be most easily understood in the limit of strong disorder, in which the variance of the random potential, $W$, is much larger than the tunnelling between neighboring sites of the lattice, $t$. In that limit, resonant transition between typical neighboring sites are impossible. The same holds for transitions between sites separated by long distances: indeed, tunnelling processes between two sites at a distance of $\sim n$ lattice sites apart typically occurs in the $n$th level in perturbation theory, and is therefore suppressed as $t_n\sim \left( t/W \right)^n$. In contrast, the typical energy mismatch of two sites, $\delta_n$, will decay only algebraically with distance $n$, $\delta_n\sim W/n^d$, where $d$ is the number of spatial dimensions. This simple argument intuitively explains, why long-range hopping processes remain off-resonant, and the wave functions are truly localized in the strong-disorder limit. Anderson localization thereby leads to the  absence of diffusion, suppressing transport (Anderson insulator).

Over almost 60 years following Anderson's original paper~\cite{Anderson58}, much progress has been achieved in understanding single-particle localization. Some prominent developments include the scaling theory of localization~\cite{ScalingTheory}, multifractality of the critical wave functions at the metal-insulator transition~\cite{Mirlin}, as well as understanding intricate effects of symmetries (e.g., time-reversal) on localization (see~\cite{50years} for a review).  Since an Anderson insulator is non interacting, it is not clear if it is a true phase of matter, and a key challenge envisioned in a pioneering work of  Anderson, which remained open for several decades, was to understand the fate of localization in the presence of interactions between particles. The new, interaction-induced processes which may potentially destroy localization are illustrated in Fig.~\ref{Fig:Anderson}c.

The interplay of interaction and disorder was addressed in the early groundbreaking work:~\textcite{Anderson80} provided qualitative arguments in favor of localization in the presence of weak short-ranged interactions. Renormalization group approaches have been used to generalize the notion of the Anderson insulator to describe the quantum ground states of interacting systems. In this context \textcite{Fink} extended the scaling theory of localization to account for the interplay of interactions and disorder. \textcite{Giamarchi88} developed a controlled renormalization group approach to describe zero-temperature properties of disordered, interacting 1d quantum liquids. 

More recently, the existence of the localized phase at {\it non-zero temperatures}, as a dynamical phase of matter, was put on a firm footing. The possibility of localization  in an interacting setting was  established for a zero-dimensional case of a quantum dot~\cite{Levitov97}, and in higher-dimensional systems with local interactions~\cite{Basko06,Mirlin05}. Such a perfect interacting insulator at non-zero temperature is said to be {\it many-body localized} (MBL). Many-body localization represents a robust dynamical phase of matter because it is stable within a range of interaction and other Hamiltonian parameters. 

We emphasize that the question whether a given interacting system is MBL is fundamentally different from the issue of the Anderson localization of its ground state. In order to establish MBL, one has to consider states with a finite density of excitations above the ground state, or, equivalently, states with a finite energy density, and show that they remain localized. In contrast, zero-temperature localization only requires the localization of a finite number of excitations in the whole system, corresponding to a vanishing energy density as the system size is taken to infinity.

From the fundamental theoretical perspective, MBL provides the only known \emph{robust} mechanism to avoid thermalization in a closed system. Other examples of systems that do not thermalize are non-interacting systems, and Yang-Baxter integrable quantum models in one spatial dimension, where any multi-particle interaction process can be reduced to two-particle collisions~\cite{Sutherland}. Unlike MBL, these are not robust with respect to small perturbations: generally adding even weak interactions or changing the form of the Hamiltonian leads to thermalization (see~\cite{Polkovnikov-rev} and references therein). Thus, such models do not describe stable phases of matter.

Recently, the phenomenon of MBL has been investigated extensively, both in theory and experiment. This led to many exciting developments and new research directions. Much of this progress, on the theory side, was fuelled by applying quantum information concepts, such as quantum entanglement, to describe the miscroscopic structure of MBL eigenstates and dynamics in those systems. Theoretical advances have largely been guided by the new experimental capabilities, which shifted the focus from traditional condensed matter setups (e.g., linear-response measurements of conductivity) to setups, which are naturally realized in isolated synthetic quantum systems (quantum quench experiments being one of the main examples). The goal of this Colloquium is to review the recent progress and current status  of MBL in an accessible manner. Below we briefly summarize  several key developments, which determine the structure and will be the main focus of this Colloquium.

We start Section~\ref{Sec:MBL} with a brief review of thermalization in quantum models. Afterwards, we introduce the notion of many-body localized phase and survey its early studies. In the main part, Section~\ref{Sec:MBL-LIOM}, we outline the \emph{phenomenological theory of MBL phase}. The key insight of this theory is that MBL systems exhibit a new kind of integrability: they are characterized by the emergence of an extensive set of quasi-local integrals of motion (LIOMs). The emergent integrability  strongly constrains the systems dynamics and thus provides an intuitive explanation of  why it fails to thermalize. We relate the \emph{entanglement structure} of MBL eigenstates to  the emergent integrability. Finally, we discuss the robustness of the emergent integrability of the MBL phase, which distinguishes it from other integrable systems.

The remainder of Section~\ref{Sec:MBL} is devoted to exploring properties of the MBL phase. Section~\ref{Sec:MBL-dyn} discusses \emph{dynamical properties} of the MBL phase that follow from a LIOM description. We explain the logarithmic growth of entanglement entropy in time -- a property which is taken as one of the key characteristics of the MBL phase. For the local observables, the entanglement spreading implies their equilibration  to highly non-thermal values at long times set by initial conditions.  Furthermore, we discuss the effect of dissipation on the dynamics of MBL systems.  Finally, Section~\ref{Sec:MBL-methods} discusses new efficient algorithms to obtain highly excited MBL eigenstates, that are possible due to the simple entanglement structure of MBL eigenstates.

In Section~\ref{Sec:MBL-enabled} we discuss the new phenomena made possible by the fact that MBL phases avoid thermalization and are not described by statistical mechanics. This allows for \emph{localization-protected quantum orders} in eigenstates, e.g. infinite temperature breaking of discrete symmetry in one-dimensional systems, which would be otherwise prohibited by statistical mechanics. In a different direction, we discuss that MBL is possible in systems with parameters that periodically depend on time (periodically driven or Floquet systems), hence preventing energy absorption and equilibration to the infinite temperature states. This makes MBL an essential ingredient that can provide the thermodynamic stability of new driven phases, such as time crystals and anomalous Floquet insulators. Such phases provide examples of new states which are only possible out-of-equilibrium. 

Section~\ref{Sec:Tran} summarizes the present understanding of the MBL \emph{delocalization transition}. This is a novel kind of dynamical phase transition between MBL and ergodic phases. From the MBL phase, the transition can be visualized as a proliferation of resonances as one increases interaction strength or decreases disorder. On the other hand, when moving from the ergodic phase,  Griffiths effects which create  bottlenecks in the transport become progressively important, especially in one dimension. We discuss the basic setup and predictions of the existing renormalization group approaches.

Section~\ref{Sec:exp} summarizes recent \emph{experimental developments} in studies of MBL. To be in the MBL phase at non-vanishing temperature, the system must be isolated from any external heat bath. In disordered solids, unavoidable coupling to a bath of delocalized phonons ultimately destroys the localized state of the electrons, leading to slow transport by variable-range hopping. However systems of,  for example, ultracold atoms are phonon-free, and thus allow for a better control of residual couplings to the environment. Thereby they offer a laboratory to observe and systematically study many-body localization and thermalization phenomena. More recently,  trapped ions, superconducting qubits, and spins of NV-centers in diamond have also emerged as promising systems where thermalization can be studied, and new non-equilibrium phases of matter can be realized.
 
 Finally, Section~\ref{Sec:outlook} concludes this Colloquium by presenting a broader perspective on the ongoing research efforts aimed to understand the quantum non-ergodic behaviors. We will outline some open questions, and  discuss future directions and possible synergies between research on MBL systems and other fields.

\section{The many-body localized phase\label{Sec:MBL}}

\subsection{Thermalization in quantum systems \label{Sec:MBL-Therm}}

We will start by discussing thermalization in isolated quantum systems. In particular, we will review the eigenstate thermalization hypothesis (ETH), which explains the microscopic mechanism of thermalization in isolated quantum systems. We will further discuss its implications for the entanglement properties of eigenstates. Since the main focus of this review is on MBL, our discussion of thermalization will inevitably be brief; a more complete overview can be found in the original papers~\cite{DeutschETH,SrednickiETH,Srednicki99} and reviews~\cite{Polkovnikov-rev,Huse-rev}. 

First, let us recall that thermalization, and more generally the statistical mechanics description of {\it classical} systems are based on the powerful \emph{ergodicity hypothesis}, which states that over a long period of time, all microstates of the system are accessed with equal probability. Directly translating this definition of ergodicity to quantum systems is problematic, since quantum mechanics operates in Hilbert space where dynamics is unitary and one cannot track a trajectory in the phase space.

To see this, let us consider an isolated quantum many-body system with a Hamiltonian $\hat H$. While the discussion below applies to general local lattice Hamiltonians (and can be further extended to continuum models),  as a concrete example the reader may keep in mind an interacting chain sketched in Fig.~\ref{Fig:1}. The generic initial non-equilibrium state $|\psi(0)\rangle$ can be expanded over the basis of many-body eigenstates $|\alpha\rangle$ as $|\psi(0)\rangle=\sum_\alpha A_{\alpha}|\alpha\rangle.$ 
Over the course of the quantum evolution, each coefficient $A_{\alpha}$ acquires a phase factor determined by the corresponding eigenenergy $E_\alpha$,
\begin{equation}\label{Eq:psit}
|\psi(t)\rangle=e^{-i\hat{H}t}|\psi(0)\rangle = \sum_\alpha A_{\alpha}e^{-iE_\alpha t}|\alpha\rangle.
\end{equation}
The probability of finding the system in a given eigenstate $|\alpha\rangle$, $p_\alpha=|A_\alpha|^2$, is set by the choice of the initial state  and does not change over time. This is unlike classical systems, which during their evolution explore different states in  phase space. Thus, we need to modify the notion of ergodicity in the quantum case. 

Intuitively, thermalization in an isolated system means that, starting from a  physical initial state\footnote{By physical we mean e.g.\ product states, extensive superposition of many eigenstates or any other states that can be experimentally prepared. In contrast, an individual eigenstate of  generic many-body system is inaccessible, as its preparation requires time which is exponentially long in system's size.} the system's observables reach values given by the microcanonical (and Gibbs) ensembles at sufficiently long times. The infinite-time average of a physical observable described by an operator $\hat O$ (which is typically a linear combination of few-body operators) can be found from (\ref{Eq:psit}):
\begin{equation}\label{Eq:O_inf}
\langle \hat O \rangle_\infty=\lim_{T\to \infty}\frac{1}{T} \int_0^T \langle \psi(t)|\hat O|\psi(t)\rangle \, dt=\sum_{\alpha} p_\alpha \langle \alpha| \hat O |\alpha\rangle. 
\end{equation}
Thus, $\langle \hat  O \rangle_\infty$ is encoded in the probabilities $p_\alpha$ along with the expectation values of the observable~$\langle\alpha| \hat O |\alpha\rangle$, because the terms that involve off-diagonal matrix elements of $\hat O$ oscillate at different frequencies and therefore average out.   Since $p_\alpha$ are fixed by the initial state, the natural way to ensure that an observable $\hat O$ reaches a thermal expectation value at long times for generic initial states is to assume that the expectation values \emph{in individual eigenstates}, $\langle\alpha| \hat O |\alpha\rangle$, agree with the microcanonical ensemble.

Such an explanation of thermalization using properties of individual eigenstates, proposed by \onlinecite{DeutschETH} and \onlinecite{SrednickiETH}, is known as the eigenstate thermalization hypothesis (ETH). 
More precisely, the ETH states that in ergodic systems, \emph{individual} many-body eigenstates have thermal observables, identical to microcanonical ensemble value at energy $E=E_\alpha$, $\langle \alpha| \hat O |\alpha\rangle\approx \mathcal{O}_\text{mc}(E)$. Thus, even if the entire system is prepared in an eigenstate, its subsystems experience the remainder as an effective heat bath, and explore possible configurations, restricted only by global conservation laws (e.g.\ energy). In this sense the ETH mechanism of thermalization implies ergodicity, so in what follows we use both notions interchangeably. ETH has been extensively tested in numerical simulations of small quantum systems~\cite{Polkovnikov-rev}.  While all known examples of thermalizing systems obey ETH, at present it is not clear if ETH is a necessary condition for thermalization.

The ETH, as formulated above, implies thermalization at infinitely long times. More specifically, since for physical initial states, the probabilities $p_\alpha$ are concentrated around a certain energy, from Eq.~(\ref{Eq:O_inf}) one can show that $\langle O \rangle_\infty\approx \mathcal{O}_\text{mc}$.   However, in order to describe the approach to the equilibrium values and bound the temporal fluctuations, one needs further information about \emph{off-diagonal} matrix elements. \onlinecite{Srednicki99} introduced the following ansatz for both diagonal and off-diagonal matrix elements of local operators $\hat O$ in the basis of eigenstates,
\begin{equation}\label{Eq:ETH-O}
\langle \alpha |\hat O |\beta \rangle = \mathcal{O}_\text{mc}(\bar E)\delta_{\alpha\beta}+ e^{-S_{\text{th}}^{\bar E}/2} R_{\alpha\beta} f(\omega,\bar E),
\end{equation}
where $\bar E  = (E_\alpha+E_\beta)/2$ denotes the average eigenenergy, and $\omega=E_\alpha-E_\beta$ is the energy difference.  $S_{\text{th}}^{\bar E}$ is the thermodynamic entropy, and $R_{\alpha\beta} $ is a normal-distributed random number. 
The expectation value of local observable and the spectral function, denoted as $\mathcal{O}_\text{mc}(\bar E)$ and $f(\omega,\bar E)$ respectively,  are smooth functions of $\omega, \bar{E}$. \textcite{Srednicki99} demonstrated that such an ansatz Eq.~(\ref{Eq:ETH-O}) is sufficient to ensure thermalization; it remains an open question whether this is also a necessary condition~\cite{Polkovnikov-rev}.

ETH has direct implications for the structure, and in particular for the entanglement properties of ergodic eigenstates. For an eigenstate $|\alpha\rangle$ obeying ETH, all observables within a sufficiently small subsystem $A$ will have thermal expectation values. This implies that the reduced density matrix of this subsystem, $\rho_A = \mathop{\rm tr}_{B} |\alpha\rangle\langle \alpha |$ (here $B$ is the complement of $A$) is thermal. Therefore, the entanglement entropy of $A$ in state $|\alpha\rangle$, which is defined as the von Neumann entropy of $\rho_A$ is equal to the thermodynamic entropy:
\begin{equation}\label{Eq:ent-ent}
 S_\text{ent}(A) =- \mathop{\rm tr} \rho_A \log \rho_A=S_{\rm th}(A).
\end{equation}
Since thermodynamic entropy is extensive, this implies that for highly excited eigenstates $|\alpha\rangle$ the entanglement entropy obeys ``volume-law'',  scaling proportionally to the volume of the subsystem, $S_\text{ent}(A)\propto \mathop{\rm vol}(A)$. As we will see below, entanglement properties of MBL eigenstates are dramatically different. 

 The matrix element ansatz Eq.~(\ref{Eq:ETH-O}) also implies the strong sensitivity of ergodic eigenstates to external perturbations of the Hamiltonian. Let us perturb the Hamiltonian by adding a small term $\epsilon \hat O$ to it, and ask how this modifies the eigenstates. 
For a typical operator $\hat{O}$, the function $f(\omega, \bar E)$ exhibits an algebraic decay for $\omega\lesssim J$, where $J$ is the characteristic energy scale of $\hat H$ (e.g.\ hopping and local interaction strength). This decay saturates at the Thouless energy~\cite{Polkovnikov-rev}. Then, Eq.~(\ref{Eq:ETH-O}) implies that the off-diagonal matrix element is exponentially larger than the many-body level spacing, which scales as $\Delta\sim J e^{-S_{\text{th}}^{\bar E}}$. Therefore, a small local perturbation of Hamiltonian, $\epsilon \hat O$ with $\epsilon\ll 1$, generally has a non-local effect in  Hilbert space, mixing an exponentially large number of original eigenstates (since $S_\text{th}^{\bar{E}}$ is proportional to the system's volume). Thus, the new eigenstates in the presence of a small perturbation are very different from the original ones, reflecting a chaotic character of ergodic quantum systems. 

Finally, the sensitivity of ETH eigenstates to external perturbations implies level repulsion. The  statistics of many-body level spacings, $s_\alpha=E_{\alpha+1}-E_{\alpha}$, where $\{E_\alpha\}$ is the ordered set of eigenenergies, was previously established to be an indicator of quantum chaos for few-body systems~\cite{Wigner}, e.g., stadium billiards  (see~\cite{Polkovnikov-rev} and references therein).   In particular, zero-dimensional quantum systems whose classical counterpart has chaotic dynamics display level repulsion. The level spacings in such systems obey Wigner-Dyson statistics, where the probability density $p(s)$ vanishes as a power-law $s^\beta$,  as $s\to 0$ with $\beta=1,2,4$ depending on the symmetry class of the model. Wigner-Dyson statistics was also found in thermalizing many-body lattice models~\cite{Polkovnikov-rev}. In contrast, if the system has an extensive number of integrals of motion (as is the case for e.g.\ Bethe-ansatz integrable models), the eigenenergies that belong to different sectors behave as independent random variables. Hence, in such systems the distribution of level spacings is Poisson, and $p(0)=1$. 

As we will demonstrate below, in the MBL phase the eigenstates break ETH and display very different properties. In particular, entanglement scaling obeys area-law (in contrast to the volume-law for ergodic systems), the effect of local perturbations remains local, and the eigenspectrum has Poisson level statistics.

\subsection{Escaping thermalization by disorder \label{Sec:MBL-Escape}}

What are the possible routes of escaping thermalization?  As we discussed in the introduction, the absence of transport would be sufficient, and therefore, Anderson localization in single-particle disordered systems provides a natural starting point to look for non-thermalizing systems. In fact, Anderson himself pointed out this possibility in 1958, stating that a localized system provides ``an example of a real physical system with an infinite number of degrees of freedom, having no obvious oversimplification, in which the approach to equilibrium is simply impossible"~\cite{Anderson58}.

For non-interacting particles on a lattice there exist two different possibilities  of localization, depending on the dimensionality and form of disorder. In low spatial dimensions ($d=1,2$) and for random uncorrelated disorder, all single particle states can be localized for arbitrarily weak disorder.  In spatial dimensions  $d\geq 3$, however, systems may exhibit a metal-insulator transition as a function of disorder strength. Such a transition is manifested by the appearance of a single-particle mobility edge in the energy spectrum, which separates localized states at low energy from extended states~\cite{50years}.  Single-particle mobility edges can also exist in lower spatial dimensions with quasiperiodic rather than random potential \cite{boers2007,xiaoli2017,lueschen2017spme}, see Section~\ref{Sec:exp}.

A fully localized Anderson insulator (without the mobility edge) of non-interacting particles is not able to transport energy or charge, and has zero conductivity at any temperature. However, in realistic systems, interactions between particles are inevitable, and therefore, to claim victory over thermalization, their effect on localization must be investigated. Interactions may open up new transport channels: in particular, a particle can decay from a high-energy (single-particle) localized state by producing an avalanche of excitations at lower energies, potentially restoring transport.  \textcite{Anderson80} related such delocalization to a non-vanishing broadening of single particle levels by interactions. Within  second order perturbation theory they showed that such broadening vanishes when disorder is sufficiently strong, suggesting the stability of localization for short-range  interacting systems. 

Later,~\textcite{Levitov97} examined interaction-induced decay of an excitation in a {\it zero-dimensional} quantum dot beyond perturbation theory. In this work, the process where a particle decays into a shower of particle-hole excitations was formulated as a single-particle Anderson localization problem in a Fock space.  Utilizing the approximate mapping to localization on a graph without loops,~\textcite{Levitov97} showed that the states below a certain energy remain localized. Despite the presence of interactions, these localized states remain close to a non-interacting single-particle excitation with a perturbatively small admixture of a few particle-hole excitations. 

Building on these results, ~\cite{Basko06,Mirlin05} analyzed the stability of the Anderson insulator with respect to short-ranged interactions.  As a starting point, they considered a model in which all single-particle states are localized, with a typical localization length $\zeta$. Interactions are characterized by a dimensionless parameter $\lambda$, given by the ratio of the two-particle transition matrix element to the level spacing $\delta_\zeta$ of excitations in a ``localization volume'' of size $\zeta$.  \textcite{Basko06} calculated the broadening of a single particle level using the self-consistent Born approximation. This approach captures the subset of decay processes where a particle at each step decays into a maximal possible number of excitations, hence maximizing the available phase space. They argued that the problem of level broadening is reminiscent of the Anderson localization problem on a tree with connectivity $K$ that depends on the temperature $T$: $K\sim {T}/{\delta_\zeta}$, which, intuitively, stems from the fact that at higher temperatures the available phase space for decay processes grows. Using an analogy with localization on a Cayley tree~\cite{Chacra}, \textcite{Basko06} estimated the critical temperature below which interacting model is localized as $K_c\approx {T_c}/{\delta_\zeta}\approx {1}/({\lambda \ln|\lambda|})$. The vanishing probability to have a non-zero decay rate of a single particle level below a critical temperature $T\leq T_c$, serves as a criterion of the stability of a  localized phase for a finite range of interaction strength in arbitrary spatial dimensions. The resulting \emph{perfect} insulator at non-vanishing temperature is termed \emph{many-body localized} phase.

At high energy densities, $T>T_c$, the phase space for allowed transitions is increasing and the many-body eigenstates in the model of \textcite{Basko07} become delocalized. The transition between localized and delocalized many-body eigenstates happens at a finite \emph{energy density}, which was dubbed a ``many-body mobility edge''~\cite{Basko07}.  The perfectly insulating behavior however, is difficult to observe in conventional solid state systems, since phonons are protected from localization and can act as a heat bath, giving rise to slow, variable-range hopping transport of localized electrons. 

An important step, which opened the door to investigating the properties of the MBL phase in numerical simulations, was taken by~\textcite{OganesyanHuse}. They pointed out that disordered lattice models with a finite dimension of the local Hilbert space can remain in the MBL phase even at infinite temperature. As a specific model, they studied a 1d chain of spinless fermions with an on-site disorder, nearest neighbor interactions, and hopping between nearest neighbor and next nearest neighbor sites. Subsequent studies concentrated on a simpler model without longer-range hopping, 
\begin{equation}\label{Eq:fermionic}
\hat H=t\sum_{i} (\hat c_{i}^\dagger \hat c_{i+1}+{h.c.}) + V\sum_i \hat n_i \hat n_{i+1}+\sum_i \epsilon_i \hat n_i, 
\end{equation}
where $\hat n_i=\hat  c_i^\dagger \hat c_i$ is the density operator on site $i$, and $\epsilon_i\in [-W;W]$ is the disorder potential distribution. Using numerical exact diagonalization of finite-size lattice systems, \textcite{OganesyanHuse} demonstrated signatures of  an MBL phase in fermionic spin chains.  

\begin{figure}
\begin{center}
\includegraphics[width=0.9\columnwidth]{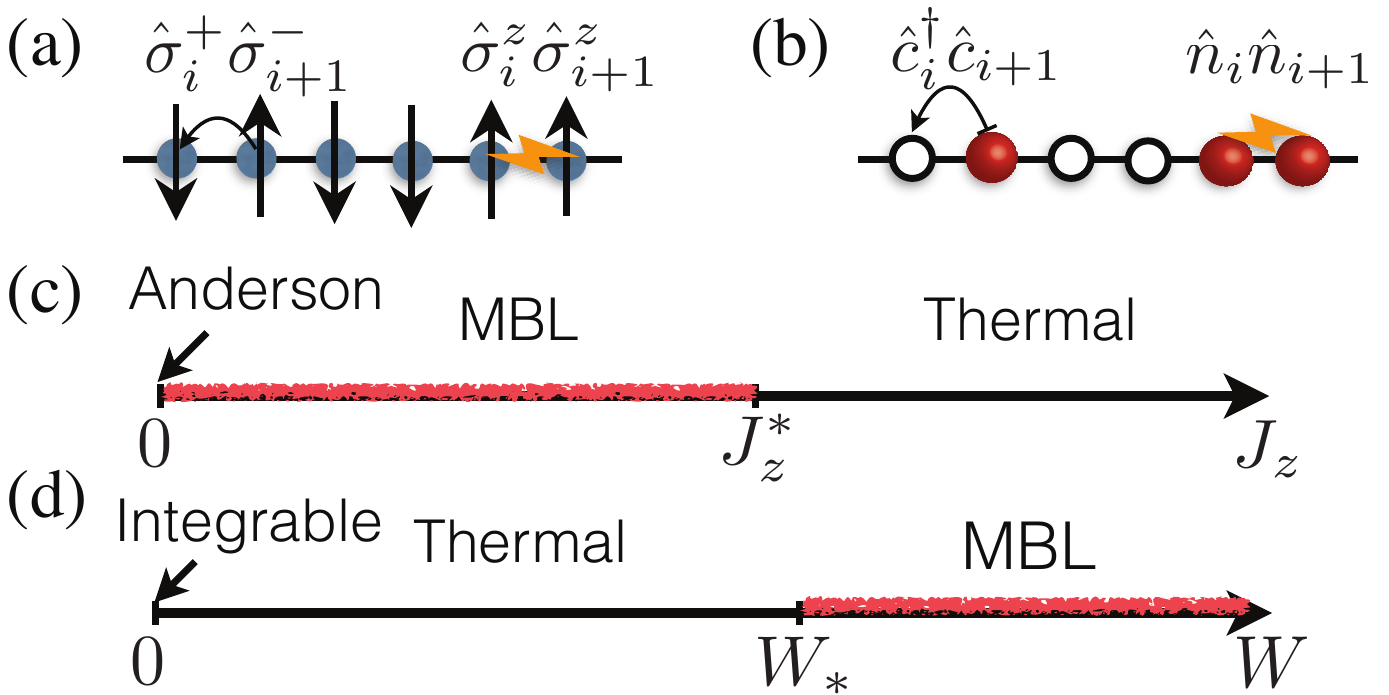}
\caption{ \label{Fig:XXZ}
Sketch of the Heisenberg spin chain (a) and spinless fermions in one dimension (b), which are used as a generic model for the MBL phase. Bottom panels show the phase diagram of the spin chain as a function of interaction (c) and disorder strength~(d). 
}
\end{center}
\end{figure}

Furthermore, \textcite{Znidaric08} and subsequently \textcite{PalHuse} presented extensive numerical studies of a disordered Heisenberg spin chain, defined by the Hamiltonian: 
\begin{equation}\label{Eq:XXZ}
\hat H_\text{XXZ}  = \frac{J_\perp}{2}  \sum_{i=1}^L (\hat\sigma_i^x \hat\sigma_{i+1}^x+ \hat\sigma_i^y \hat\sigma_{i+1}^y)+\frac{J_z}{2}\sum_{i=1}^L \hat\sigma_i^z \hat\sigma_{i+1}^z + h^z_i \hat\sigma^z_i,
\end{equation}
where $\hat{\bm \sigma}= (\hat\sigma^x,\hat\sigma^y,\hat\sigma^z)$ is a vector of three spin-1/2 Pauli operators, and $h^z_i$ are randomly distributed on-site magnetic fields, $h^z_i\in [-W,W]$ (thus, $W$ is the disorder strength).  As illustrated in Fig.~\ref{Fig:XXZ},  this random-field XXZ spin chain can be mapped onto the  chain of spinless fermions in Eq.~(\ref{Eq:fermionic}), using a Jordan-Wigner transformation~\cite{JordanWigner}. In this mapping, $J_\perp$ and $J_z$ terms are transformed into fermion hopping and nearest-neighbor interaction, respectively, while the $h_i^z \hat\sigma_i^z$ term is transformed into random on-site energies. 

In the limit $J_z\to 0$, the  XXZ spin chain  is equivalent to free fermions moving in a disorder potential, and therefore, all states are Anderson localized, for arbitrary values of disorder strength $W$.  For a fixed and not too large $W$ (i.e. $W\sim 1$) all many-body states remain localized as long as the interaction is below some threshold value $|J_z|<J_z^*(W)$. The corresponding schematic phase diagram is sketched in Fig.~\ref{Fig:XXZ}(c), where we do not illustrate a possible recurrent MBL phase for very strong $J_z\gg J_\perp$. Similarly, at fixed interaction strength there is a critical value $W_*$ above which the many-body states become fully localized, see Fig.~\ref{Fig:XXZ}~(d). The numerical results~\cite{Alet14, Serbyn15} also indicate that for disorder $W<W_*$, eigenstates in the middle of the band become delocalized leading to a many-body mobility edge. Thus, already in one dimension there is a transition between a localized and a thermal phase -- a property which distinguishes MBL from single-particle Anderson localization.

In the localized regime  \textcite{PalHuse} observed a breakdown of ETH through a number of metrics. In particular, the spin expectation value $\bra{\alpha} \hat\sigma^z_i\ket{\alpha}$ in eigenstates was found to fluctuate widely between adjacent many-body eigenstates, in contradiction to the ansatz Eq.~(\ref{Eq:ETH-O}), requiring local observables to be smooth functions of energy. To locate the critical disorder strength for which the system enters  an MBL phase, Refs.~\cite{OganesyanHuse,PalHuse} used the average ratio of adjacent level spacings as an diagnostic probe of the level statistics. \textcite{PalHuse} confirmed that for weak disorder there is level repulsion and the level statistics is of Wigner-Dyson form,  as expected for a thermalizing system,  see Section~\ref{Sec:MBL-Therm}. In contrast, for strong disorder the level repulsion disappears and the level statistics approaches a Poisson distribution, which, as we will explain in the next section, is a consequence of a new form of  emergent integrability present in the MBL phase. For $J_z=J_\perp=1$ the crossover point between the two behaviors, $W_*\approx 3.5$ was identified as the location of  the transition between MBL and ergodic phases.

The introduction of such microscopic lattice models enabled investigations of highly non-equilibrium dynamics of  the MBL phase, as opposed to traditional computations of Ohmic conductivity. \onlinecite{Znidaric08} and \onlinecite{Moore12} numerically studied the behavior of a model of Eq.~(\ref{Eq:fermionic}) in an MBL phase in a quantum quench experiment, shown in Fig.~\ref{Fig:1}. First, they numerically observed that starting from initial product states, while there was no transport, entanglement between two parts of the spin chain kept growing logarithmically in time. Such a growth was absent in the Anderson insulating phase.  The second surprising observation was that at very long times, the entanglement entropy saturated at values which were proportional to the system size, albeit smaller than the expected thermal value of the entropy at the same energy density. This result demonstrated that quantum information spreads throughout the entire system and indicated that the MBL phase has qualitatively different properties compared to the non-interacting Anderson insulator.  

Below we introduce an effective theory of the MBL phase based on local integrals of motion, which provides a unified description of most known properties of the MBL phase, such as absence of transport, integrability, logarithmic dynamics of entanglement, as well as  the breakdown of ETH.  In addition, we  discuss other predictions of this theory, in particular, the dynamics of local observables in a quantum quench experiment.  

\subsection{Emergent integrability of MBL phase \label{Sec:MBL-LIOM}}

In this Section we introduce a new kind of integrability, which characterizes the MBL phase. The construction builds on the entanglement structure of the eigenstates in this phase.  We start with a heuristic argument, which exploits the intuitive definition of MBL (in the absence of a many-body mobility edge), as a phase where local perturbations have only local effects on the eigenstates. This intuition can be used to understand the entanglement structure of eigenstates and to infer the existence of local integrals of motion. In addition, we discuss an alternate viewpoint on how the local integrals of motion emerge.  Finally, we demonstrate how integrability explains breakdown of thermalization, and contrast MBL systems to other examples of integrable systems. 

\subsubsection{Area law entanglement in MBL eigenstates}

MBL eigenstates display a low amount of entanglement, obeying the so called area-law: that is, the entanglement entropy of a subsystem $A$ in an MBL egeinstate scales proportional to the volume of the boundary $\partial A$ of $A$, as both the size of the system and the size of $A$ are taken to infinity, $S_{\rm ent} (A)\propto {\rm vol}(\partial A)$. Area-law entanglement scaling is typical of ground states in gapped systems~\cite{VerstrateReview,EisertAreaLaws}. However, as we explain below, in MBL systems even highly excited states obey area-law scaling, in contrast to thermal eigenstates which have volume-law entanglement. 

\begin{figure}[t]
\begin{center}
\includegraphics[width=0.8\columnwidth]{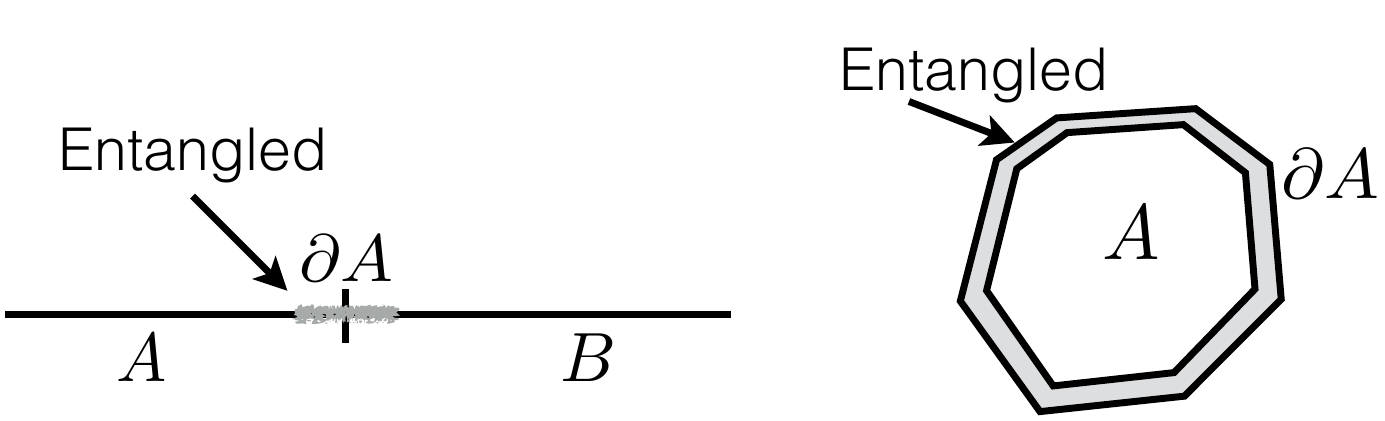}\\
\caption{ \label{Fig:ent-structure}
 Illustration of the area-law entanglement entropy in one and two spatial dimensions where only the shaded boundary regions $
 \propto\partial A$ contribute to the entanglement. In contrast, for systems with volume-law entanglement, extensively many degrees of freedom  $\propto {\rm vol}(A)$ are entangled with the exterior region.}
\end{center}
\end{figure}

The low entanglement of MBL eigenstates can be intuitively inferred from the following thought experiment~\cite{Serbyn13-1}. Let us consider an MBL system with a local Hamiltonian $\hat  H$, and specify a region $A$ (e.g., a block of adjacent spins in a one-dimensional spin chain, in which case $\partial A$ is just two end points of the block). We divide the Hamiltonian into three parts: $\hat H_A$, which contains the terms acting only on spins in $A$, $\hat  H_B$ acting only on spins in the complement of $A$, and terms $\hat  V_{AB}$ which couple spins in $A, B$ along the boundary $\partial A$. Let us turn off the couplings along the boundary of region $A$. Then, the eigenstates are simply tensor product states of eigenstates $|\alpha\rangle_A$, $|\beta\rangle_B$ of $\hat  H_A$ and $\hat  H_B$:
\begin{equation}\label{Eq:product}
|I\rangle_{AB}=|\alpha\rangle_A \otimes |\beta\rangle_B. 
\end{equation}
These states have zero entanglement entropy for region~$A$. Now, let us turn on the coupling $\hat V_{AB}$, which acts locally near the boundary. Since the system is in the MBL phase, introducing a local perturbation will only significantly affect degrees of freedom situated within the localization length $\xiAB$ from the boundary.\footnote{We will provide a more precise definition of the many-body localization length below.} Thus, we expect that the new eigenstates can be obtained from the states $|I\rangle_{AB}$ by entangling spins in $A$ and $B$ over a distance $\sim \xiAB$ away from the boundary $\partial A$. The effect of introducing a local perturbation on spins far away from the boundary is expected to decay exponentially with the distance leading to an area-law scaling of entanglement entropy $ S_{\rm ent}\propto {\rm vol} (\partial A)$. The area-law entanglement scaling of MBL eigenstates, suggested by this argument, was demonstrated numerically in~\cite{Serbyn13-1,Bauer13}. As we discuss in Section~\ref{Sec:MBL-dyn}, despite the area-law entanglement of eigenstates, the dynamics of the MBL phase after global quench leads to a volume-law saturation value of entanglement~\cite{Serbyn13-1}.

\subsubsection{Quasi-local integrals of motion}
The low entanglement of MBL eigenstates implies that they can be connected to product states by a sequence of quasi-local unitary transformations~\cite{Serbyn13-1} except for the case when MBL eigenstates exhibit topological order~(see Section~\ref{sec:SPT}).  Such unitary transformations diagonalize the Hamiltonian in a given product state basis. Their quasi-local nature can be used to map physical degrees of freedom into \emph{quasi-local} integrals of motion. 

To make this intuition more precise,  let us consider the disordered Heisenberg model of Eq.~(\ref{Eq:XXZ}). In the limit $J_\perp\to 0$ the Hamiltonian $\hat  H_0=\sum h_i \hat \sigma_i^z+J_z\sum \hat \sigma_i^z\hat \sigma_{i+1}^z$ commutes with the $\hat \sigma_i^z$ operator on every site, and therefore the eigenstates are non-entangled product states, where each spin has a definite $z$-projection: 
\begin{equation}\label{Eq:product_Z}
|\{\sigma \}\rangle=|\sigma_1\sigma_2\dots \sigma_N\rangle, \,\, \sigma_i=\uparrow,\downarrow. 
\end{equation}
In total, we have $2^L$ eigenstates,  where $L$ is the number of spins, labeled by strings $\{\sigma \}$. 

\begin{figure}[b]
\begin{center}
\includegraphics[width=0.95\columnwidth]{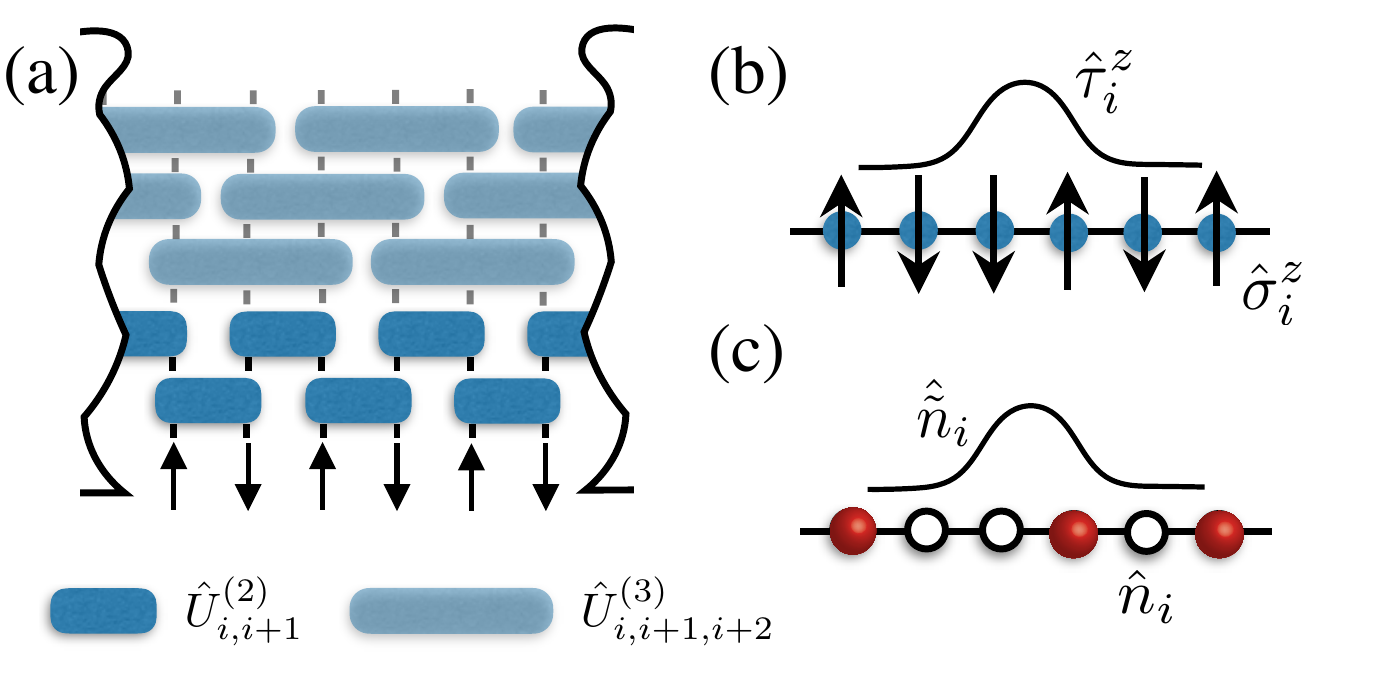}\\
\caption{ \label{Fig:LIOM} (a) Rotation of the product states into the exact many-body eigenstates can be achieved by a sequence of quasi-local unitary transformations. (b,c) The same quasi-local unitary transformation can be used  to obtain the quasi-local operators $\hat \tau^z$ and $\hat{\tilde n}_i$ which commute with the Hamiltonian.
}
\end{center}
\end{figure}

Now, let us turn on a weak flip-flop (kinetic) term $J_\perp$, such that the system remains in the MBL phase but the Hamiltonian is no longer diagonal in the $|\{\sigma \}\rangle$ basis. The above argument for the area-law entanglement implies that the new eigenstates can be obtained from the product states Eq.~(\ref{Eq:product_Z}) by a quasi-local unitary transformation. We say that $\hat U$ is quasi-local if it can be factored into a sequence of 2-site, 3-site, 4-site \ldots, unitary operators as  ${\hat  U} =  \prod_{i}  \ldots \hat{U}^{(3)}_{i,i+1,i+2} \hat{U}^{(2)}_{i,i+1} $ (see Fig.~\ref{Fig:LIOM} for a schematic illustration).  In this expansion, the long-range unitary operators have progressively decreasing rotation angles, so that   $||1- \hat U^{(n)}_{i,i+1,\ldots i+n}||^2_F < e^{-n/\xiU}$, where $||\cdot ||_F$ is the Frobenius operator norm.   In contrast, if the Hamiltonian describes a thermalizing phase, the operator $\hat U$ that diagonalizes it is  highly non-local since it rotates the product states into states with volume-law entanglement. 

The unitary operator ${\hat U}$ transforms the integrals of motion $\hat \sigma_i^z$ of $\hat H_0$ into the integrals of motion $\hat \tau^z_i={\hat U}^\dagger \hat \sigma^z_i {\hat U}$ of~$\hat H$. Because ${\hat U}$ is quasi-local in an MBL system, the $\hat \tau^z_i$ are typically close to the microscopic spin operators $\hat \sigma^z_i$,  at least at strong disorder. Specifically, $\hat \tau^z_i$ can be expanded as
\be \label{Eq:OPE}
\hat  \tau^z_i=Z\hat \sigma^z_i+ \sum_{n=1}^\infty V^{(n)}_{i}{\hat O^{(n)}}_i,
\ee
where $\hat O^{(n)}_i$ contains up to $2n+1$-body operators with contributions from sites at distance $n$ from $i$~(i.e.\ sites $ i-n,\ldots,i,\ldots i+n$ could contribute)  and is normalized to $||\hat O_i^{(n)} ||_F=1$.   Each $\hat \tau^z_i$ has a finite overlap $Z$ with the  microscopic spin operator~$\hat \sigma^z_i$. Moreover, the coefficients of longer-range operators decay as $V^{(n)}_i\sim e^{-n/\xiOPE}$, so the effect of operator ${\hat \tau}_i^z$ on spin $j$ situated far away from site $i$, is exponentially small.  This locality is the key property which distinguishes the MBL phase from the thermal phase. The lengthscale $\xi$, which controls the locality of ${\hat \tau}_i^z$, or equivalently, the locality of the unitary $\hat U$  can be viewed as the localization length in the MBL phase.

The  operators $\hat \tau_i^z$ form a complete set of independent \emph{quasi-local integrals of motion}, which we abbreviate as LIOMs in what follows (in the literature they are also sometimes called localized bits, or l-bits). Eigenstates of $\hat H$ can be fully specified by labeling them with the eigenvalues of all $\hat \tau^z_i$. One can view each $\hat \tau_i^z$ operator as an emergent conserved pseudospin-like degree of freedom; it cannot decay during quantum evolution as long as an MBL system is not coupled to an external heat bath. In principle, one could define operators ${\hat \tau}_i^z$ for thermal systems, but in that case they would be highly non-local, have vanishing overlap with the microscopic spin operators and thus they would be of little use. 

To form a complete basis of operators, we introduce 
operators $\hat \tau_i^{x,y}=\hat U \hat \sigma_i^{x,y} \hat U^\dagger$, which are also quasi-local. The operators $\hat \tau_i^{x,y,z}$ and their products form a complete basis in the operator space. Therefore, any physical operator described by $\hat \sigma_i^\alpha$ and their products can be decomposed in the $\tau$-basis.  

The $\tau$-representation is particularly useful when analyzing the dynamics in the MBL phase, as we will see below. This is due to the simple form that the system's Hamiltonian
$\hat H$ takes in the $\tau$-basis. Indeed, since $[\hat \tau_i^z,\hat H]=0$, $\hat H$ cannot include any $\hat \tau^{x,y}_i$ operators. This results in the following general form~\cite{Serbyn13-1,Huse13}:
\begin{equation}\label{Eq:H-universal}
\hat  H_\text{MBL} = \sum_{i} \tilde{h}_i \hat \tau^z_i + \sum_{i>j} J_{ij} \hat \tau^z_i\hat \tau^z_j
+ \sum_{i>j>k} J_{ijk}\hat  \tau^z_i\hat \tau^z_j\hat \tau^z_k+\ldots.
\end{equation}
Here and in what follows we denote by $\hat H_\text{MBL}$ the Hamiltonian written in terms of  LIOMs.
Now, since the above representation results from the action of the quasi-local transformation $\hat U$ on the local Hamiltonian $\hat H$, the couplings 
\begin{equation}\label{Eq:J-exp}
J_{ij}\propto J_0 e^{-{|i-j|}/{\xiJ}},
\quad
J_{ijk}\propto J_0 e^{-{|i-k|}/{\xiJ}}, \ldots
\end{equation}
decay exponentially with separation between the LIOMs. The above form of the Hamiltonian is often viewed as the universal Hamiltonian of the MBL phase. 

Assuming that the operators $\hat O^{(i)}_n$ entering Eq.~(\ref{Eq:OPE}) are superposition of Pauli strings with coefficients that follow a narrow distribution, the corresponding lengthscale $\xiJ$ can be shown to satisfy $\kappa^{-1}\geq (\xi^{-1}+\ln 2)/2$.~\footnote{To derive this inequality, we equate $\hat H_\text{MBL}$ to the original Hamiltonian, e.g.~Eq.~(\ref{Eq:XXZ}) and substitute the expansion of $\tau^\alpha_i$ via $\sigma^{\alpha}_i$ operators,  Eq.~(\ref{Eq:OPE}). The extra term $\ln 2$ accounts for the exponentially large number of possible couplings that contribute to the operator $\hat O_n^{(i)}$ of a given range $n$, and the overall factor $1/2$ accounts for their random signs.} 
This implies that $\xiJ$  must remain finite even if $\xi$ diverges at the MBL transition (see Section~\ref{Sec:Tran}). We note that it would be interesting to test both the distribution of coefficients, and the resulting bound, which so far remains a hypothesis.

It is interesting to consider the analogy between the effective MBL Hamiltonian Eq.~(\ref{Eq:H-universal}) and Landau's Fermi-liquid theory of interacting fermion systems. Within the equivalent fermion description of the XXZ chain, Eq.~(\ref{Eq:fermionic}), the operators $\hat\sigma^z_i$ are the site occupation numbers $\hat n_i$, while the  LIOMs $\hat \tau^z_i$ map to ``quasiparticle'' occupation numbers~$\hat {\tilde n}_i$. In both cases the effective theory can be written entirely in terms of the commuting integrals of motion. Furthermore, there is a non-vanishing overlap between the bare fermion or spin operators and the dressed operators~\cite{Bera15}. The main difference between the Fermi liquid and the MBL phase is that in Fermi liquids the effective theory is only valid in the low energy limit, while the MBL Hamiltonian provides an exact description  at all energies. In the former case, quasi-particle operators are true integrals of motion only for wave-vectors   asymptotically close to the Fermi surface, while in the latter case the LIOMs are a complete set of commuting operators that fully specify all eigenstates.
 
The emergent integrability in the MBL phase naturally explains the Poisson level statistics observed in the early numerical studies~\cite{OganesyanHuse,PalHuse}. It also explains the breakdown of ergodicity in dynamics because it implies that during its evolution, an MBL system retains the local memory of the initial states, encoded in the initial values of LIOMs. Moreover, in many cases the LIOMS have an overlap with conserved densities such as energy or particle number, which explains the absence of transport in the MBL phase.  We note, however, that the existence of global conserved quantities is not essential for the MBL phase -- for example, as discussed in Section~\ref{subsec:MBL-Floquet}, MBL is possible in periodically driven systems where even energy is not conserved. 

So far, we  outlined the description of~\cite{Serbyn13-1,Huse13},  which introduced the Hamiltonian in Eq.~(\ref{Eq:H-universal}) as an effective model of the MBL phase using an intuitive definition of this phase and its entanglement properties. This is reminiscent of Landau's hypothesis of the Fermi-liquid Hamiltonian based on adiabatic continuity. However,  in a modern perspective, the integrable Fermi-liquid Hamiltonian can be obtained systematically, as a renormalization group~(RG) fixed point~\cite{Polchinski1992,Shankar1993}. Similarly, the integrability of the MBL state as well as an approximate form of the LIOMS and the effective Hamiltonian were also obtained through a perturbative RG approach \cite{Vosk13,Pekker14}. 

The RG transformation can be formulated as a dynamical scheme that captures the time evolution by successively integrating out the fastest modes \cite{Vosk13}. {A related scheme}, so-called real space RG for excited states (RSRG-X) uses an approximate succession of local unitary transformations to construct eigenstates of the system at all energies \cite{Pekker14}. In the renormalization process one successively integrates out local degrees of freedom with the highest frequency scale in analogy to the strong disorder real space RG scheme of \textcite{DasguptaMa}. A crucial difference however is that in the standard scheme \cite{DasguptaMa} the eliminated degrees of freedom are always put into their lowest energy configuration, while in the RSRG-X one chooses either low-energy or the high-energy manifold of the local term. For instance, the spin with the magnetic field that is the largest scale in the Hamiltonian, $h_z\hat \sigma^z_i$, can be put into  either of configurations $\sigma^z=\pm 1$. These degrees are therefore not really eliminated, but rather become the emergent LIOMs~$\hat \tau^z_i$. By choosing the ground state/excited state of the decimated pair of spins one assigns a given value $\tau^z_i=\pm 1$ to this integral of motion. Hence in this way one is able to obtain the entire spectrum of the many-body Hamiltonian. The universal Hamiltonian Eq.~(\ref{Eq:H-universal}) emerges as a fixed point of such an RG flow.  It should be noted that the RSRG-X method is only approximate, as it only keeps track of a certain subset of many-body processes  does not capture (rare) long-range resonances that are related to the absence of the adiabatic limit in the MBL phase.

The quasi-local nature of the unitary transformation $\hat U$ and therefore the existence of the MBL phase were subsequently proven by~\textcite{Imbrie16,ImbriePRL} for the strongly disordered Ising spin chain with transverse and longitudinal magnetic fields. In essence, the strategy is to perform a sequence of more and more non-local unitary transformations which gradually diagonalize the Hamiltonian. The proof relies on a reasonable assumption that limits the attraction between many-body eigenenergies, and puts bounds on the probability to have long-range resonances which could potentially destroy the quasi-local structure of the unitary~$\hat U$. The proof makes explicit use of the one-dimensional nature of the system, thus it does not apply to higher  spatial dimensions.

Other perturbative approaches were used to obtain an approximate construction of LIOMs that would be valid also in higher dimensions.  \textcite{ScardicchioLIOM} constructed the LIOMs using a {perturbative} technique similar to the self-consistent Born approximation of~\cite{Basko06}. Localization length and other properties of LIOMs constructed {numerically} were further studied in~\cite{Chandran14,Rademaker16,Brien16,Schiro18,Gil17}.

The description of the MBL phase in terms of LIOMs is expected to be valid throughout the phase. In Section~\ref{Sec:MBL-dyn} we will explore the  physical implications of the universal MBL Hamiltonian for dynamics in the MBL phase. Upon approaching the delocalization transition, the LIOM operators are expected to become increasingly non-local due to long range resonances. Thus perturbative approaches fail to describe the delocalization transition as well as critical phenomena associated with the approach to this transition. We shall review various aspects of the delocalization transition in Section~\ref{Sec:Tran}. 

\subsubsection{Comparison to other integrable systems}

The existence of an extensive set of local conserved operators seemingly places MBL systems in the same category with other integrable models. However, as we discuss below, the MBL integrability is conceptually different compared to other previously known kinds of integrability in non-interacting systems, and  in Yang-Baxter integrable systems~\cite{Sutherland}.

First, the integrals of motion in the MBL phase are quasi-local operators,  in contrast to the integrals of motion for Yang-Baxter-integrable systems, which are extensive sums of local operators~\cite{Sutherland}. Second, the emergent integrability of the MBL phase is robust: if an MBL Hamiltonian is perturbed by a weak, but finite perturbation, the system stays in the MBL phase, and therefore a deformed set of LIOMs can be defined. In contrast, if a non-interacting system of (delocalized) fermions, characterized by conserved occupations of single-particle eigenstates, is perturbed by introducing an arbitrarily weak two-body interaction, the integrability is immediately destroyed. A similar scenario is expected to hold for Yang-Baxter-integrable systems~\cite{Polkovnikov-rev}. The robustness of LIOMs reflects the fact that MBL is a dynamical phase of matter, while non-interacting systems and Yang-Baxter integrability represent isolated points or lines in the phase space of possible Hamiltonians. 

It is also instructive to draw a parallel  between MBL and weakly perturbed integrable classical systems. For the latter, the powerful Kolmogorov-Arnold Moser~(KAM) theory, which establishes that weak integrability-breaking perturbations transform most periodic orbits into quasi-periodic ones~\cite{Arnold}. An important assumption of KAM theory is the incommensurability of frequencies, which ensures the absence of resonant processes between different degrees of freedom. Similarly, in MBL systems, the incommensurability of frequencies/energies arises naturally due to disorder. The stability of the MBL phase, where weak local perturbations deform but do not destroy LIOMs, may be viewed as the analogue of the KAM theorem for quantum many-body systems. Moreover, the MBL phase is the only known example of a KAM type integrable system that survives in the thermodynamic limit.

\subsection{Dynamical properties of MBL phase \label{Sec:MBL-dyn}}

 The emergent integrability underlies ergodicity breaking in the MBL phase, and strongly constrains the dynamics therein. In this Section, we discuss largely universal dynamical properties  that stem from the existence of LIOMs. In particular, we will analyze the behavior of an isolated MBL system following a quantum quench, and explain the origin of the logarithmic spreading of entanglement. Surprisingly, despite localization, physical observables (such as local magnetization) do equilibrate at long times, albeit to highly non-thermal values, which carry information about the initial state. This, and other dynamical signatures, which have been predicted from the theory based on LIOMs, can be understood using the effective Hamiltonian Eq.~(\ref{Eq:H-universal}) of the MBL phase.

 To illustrate the origin of entanglement generation in a localized system, let us consider the evolution of an MBL spin chain starting from a low-entanglement state.  In numerical simulations~\cite{Znidaric08,Moore12}, the initial state was typically taken to be a product state of physical spins.  For simplicity, let us instead focus on a different state: a product state in the basis of LIOMs, in which each $\tau$-spin is pointing in some direction on the Bloch sphere:
\begin{equation}\label{Eq:psi-tau}
|\psi_0\rangle= \otimes_i (A_{i}|\Uparrow_i\rangle+B_{i}|\Downarrow_i\rangle),
\end{equation}
where we introduced double arrows $\Uparrow_i,\Downarrow_i$, which refer to eigenstates of $\tau^z_i=\pm1$, and $|A_i|^2+|B_i|^2=1$. Otherwise, the coefficients $A_i, B_i$ can be arbitrary.~\footnote{One should however avoid the situation where all $A_i=1$ or $B_i=1$, because this would correspond to an eigenstate, which would take an exponentially long time (of the order of the inverse level spacing) to prepare.}

\begin{figure}[t]
\begin{center}
\includegraphics[width=0.99\columnwidth]{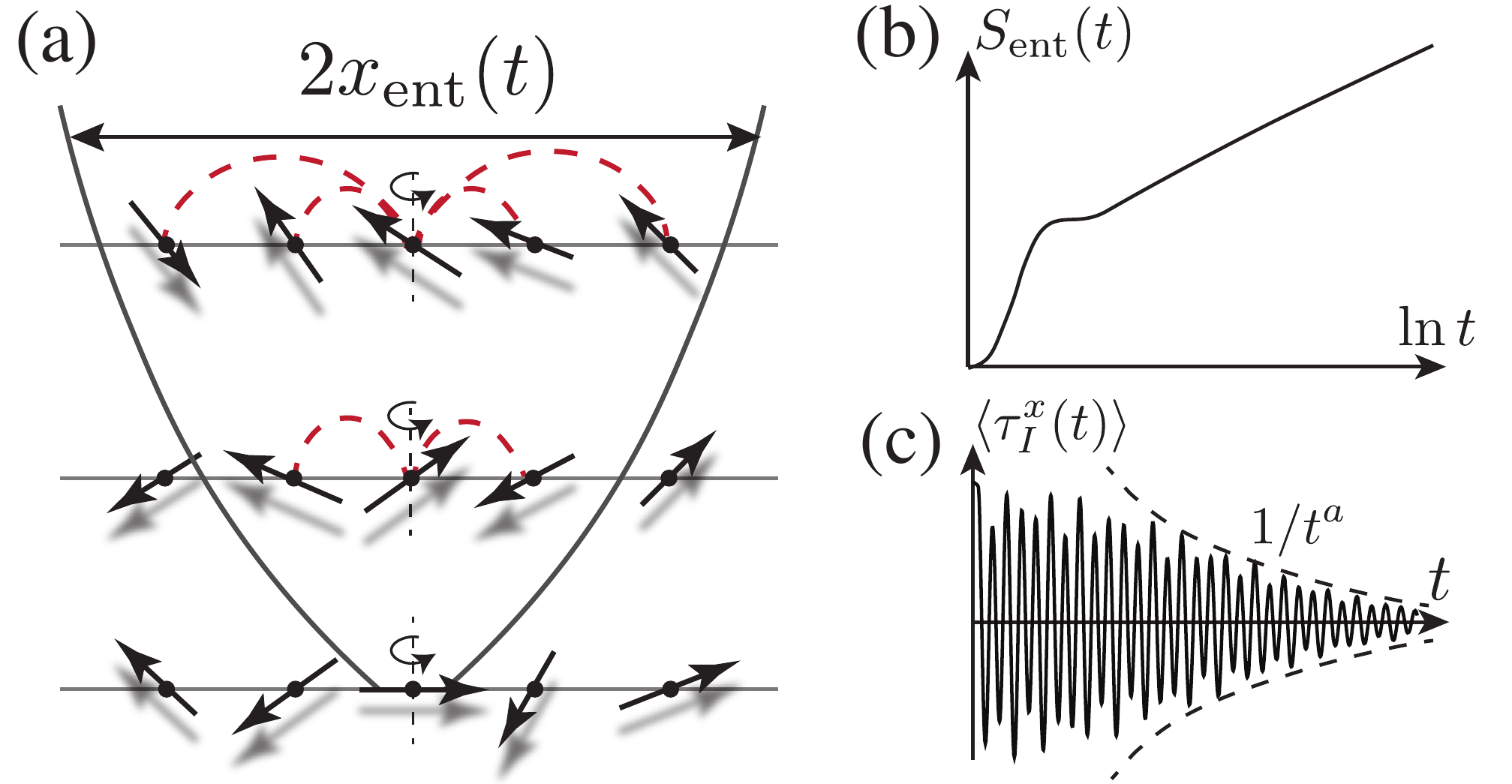}
\caption{ \label{Fig:ent-growth}
Mechanism of dynamics in the MBL phase: (a) The central LIOM precesses with time, and interactions with other LIOMS lead to the dependence of its precession frequency on states of neighboring spins. This process is responsible for logarithmic growth of entanglement shown in panel (b) and also for the relaxation of fluctuations of $\langle\tau^x(t)\rangle$ illustrated in panel (c).}
\end{center}
\end{figure}

The evolution of this state can be understood using the Hamiltonian Eq.~(\ref{Eq:H-universal}): each $\tau$-spin precesses in the external magnetic field created by other $\tau$-spins. The coefficients $A_i,B_i$ thereby acquire phases. Crucially, these phases depend on the state of other $\tau$-spins, which leads to entanglement generation, see Fig.~\ref{Fig:ent-growth}(a). 
This can be illustrated for just two LIOMs prepared in a superposition $|\psi_2\rangle= (|\Uparrow_1\rangle+|\Downarrow_1\rangle )\otimes (|\Uparrow_2\rangle+|\Downarrow_2\rangle )/2$. While initially $|\psi_2\rangle$ is a product state,  evolution with effective Hamiltonian $\hat{H}_\text{MBL} = J_{12}\hat \tau^z_1\hat \tau^z_2$ (we omit single-spin terms as they do not produce entanglement) introduces the phases that depend on the relative state of spins 1 and 2, resulting in the evolution $|\psi_2(t)\rangle =e^{-i\hat{H}_\text{MBL}t}|\psi_2\rangle=e^{-iJ_{12}t}(|\Uparrow\Uparrow\rangle+|\Downarrow\Downarrow\rangle)/2+e^{iJ_{12}t}(|\Uparrow\Downarrow\rangle+|\Downarrow\Uparrow\rangle)/2$. This wave function $|\psi_2(t)\rangle $ now has entanglement between two spins that grows up to $\ln 2$ for times when the accumulated phase $J_{12} t=\pi/4$~\cite{Serbyn13-2}. 

We sketch the generalization of this argument for the case of many spins, following~\cite{Serbyn13-2,Huse13,Serbyn13-1}. A given spin acquires phase of order one dependent on the state of another spin a distance $x$ away after a time $t(x)$ set by condition $ \tilde h_{i,i+x}t(x) \sim 1$.  Here we introduced the strength of an effective magnetic field acting on spin at site $i$ due to spin distance $x$ away. This field depends on the state of other spins between spin $i$ and $i+x$, $\tilde h_{i,i+x} = J_{i,i+x}+J_{i,i+1,i+x}\hat \tau_{i+1}^z+\ldots$. Using exponential decay of couplings $J_{ij},\ldots$, see Eq.~(\ref{Eq:J-exp}), one can show that the effective magnetic field is also exponentially suppressed with the interaction range, $\tilde h_{i,i+x}\sim J_0 e^{-x/\xiE}$, where the lengthscale $\xiE$ is discussed below. This decay of the magnetic field together with the condition $ \tilde h_{i,i+x}t(x) \sim 1$ yields  the logartihmic entanglement ``light cone" and  logarithmic growth of entanglement,
\begin{equation}\label{Eq:log-growth}
 x_\text{ent}(t) = \xiE \log (J_0 t),
 \quad 
 S_{\rm ent}(t)\propto \xiE \log(J_0t).
\end{equation}
Indeed, at time $t$, for typical initial states, all spins within volume $x_{\text{ent}}(t)$ acquire phases dependent on the states of other spins, and therefore entanglement spreads over that volume. In a finite-size system, $ x_\text{ent}(t)$ is bounded by the system size, hence the entanglement entropy in a quantum quench saturates to a value that is proportional to the system size,  $S_{\rm ent}(\infty)\propto L$.

 The length $\xiE$ which controls entanglement growth is related to $\kappa$ as $\xiE^{-1}\leq \xiJ^{-1}+(\ln2)/2$, where extra contribution comes from the exponentially large number of possible interaction terms within a given range. Thus, $\xiE$ can be shown to satisfy $\xi'\leq 2\xi$. Note, that we started with a single localization length $\xi$ in Section~\ref{Sec:MBL-LIOM}, and introduced two additional lengths $\kappa$ and $\xi'$ which control different physical properties. These lengthscales may be viewed as phenomenological parameters which could in principle account for the presence of multiple intrinsic lengthscales within MBL phase. Establishing whether  the three lengthscales $\xi$, $\xiJ$, and $\xiE$ are directly related  remains an outstanding challenge. 

Two comments are in order. First, while for simplicity we have focused on initial product states of $\tau$-spins, the logarithmic growth of entanglement holds generally, e.g., for initial product states of physical spins~\cite{Serbyn13-2}. This is because a generic initial state is an extensive superposition of many-body eigenstates, and each $\tau$-spin undergoes the dephasing dynamics. Second, the proportionality coefficient in Eq.~(\ref{Eq:log-growth}) depends on the diagonal entropy of the initial state~\cite{Polkovnikov}, which in particular is influenced by the disorder strength, as discussed in Ref.~\cite{Serbyn13-2}. For initial states of Eq.~(\ref{Eq:psi-tau}) this entropy is determined by the probability distribution of coefficients $A_i,B_i$. 

Entanglement spreading in MBL systems can also be described by the dynamical RG approach introduced by~\cite{Vosk13}. This approach gives an entanglement entropy that grows logarithmically in time for usual MBL states in agreement with Eq.~(\ref{Eq:log-growth}), while  in the vicinity of a transition between two distinct MBL phases, the growth changes to  $S_\text{ent}(t)\propto \ln^{\phi} t$ with $\phi>1$  (see Section~\ref{Sec:MBL-enabled}).

The logarithmic propagation of entanglement in the MBL phase is in a stark contrast to the ballistic entanglement spreading in ergodic systems~\cite{Kim13}  and in Yang-Baxter-integrable models~\cite{CalabreseCardyReview}. The logarithmic spreading of entanglement also distinguishes the MBL phase from an Anderson insulator. It is therefore often viewed as one of the defining features of MBL, especially in numerical simulations.

The growth of entanglement is difficult to measure experimentally in large systems (see, however, Section~\ref{Sec:exp}). It is therefore important to identify alternative manifestations of the dephasing dynamics in the MBL phase that are more directly observable. \textcite{Serbyn14} showed that the dephasing dynamics in MBL systems leads to equilibration of local observables in a quantum quench setup, with a characteristic, power-law approach to equilibrium values. 

To illustrate this, let us again consider the simple initial state (\ref{Eq:psi-tau}) with $A_i=B_i=1/\sqrt{2}$ and study the single-spin observables, described by operators $\hat \tau_I^{x,y}$ (operator $\hat \tau_I^z$, being conserved, is time-independent). At $t=0$, the $\tau$-spin $I$ is pointing in the $x$ direction. Further, it undergoes precession in the $xy$-plane, and its rotation angle depends on the state of other spins. At time $t$, spins within distance $x_\text{ent}(t)$ away, such that $tJ_0e^{-x_\text{ent}(t)/\xiE}\gtrsim 1$, significantly affect the rotation angle of spin $I$. The generation of entanglement with those spins leads to dephasing. Accordingly, the off-diagonal elements of the reduced density matrix (or, equivalently, observables $\langle \hat \tau_I^{x,y}(t)\rangle$ decay. Observing that at time $t$ spin $I$ is entangled with an ``environment" that has the Hilbert space dimension ${\cal D}(t)\approx 2^{2x_\text{ent}(t)}$ spins,  and using Eq.~(\ref{Eq:log-growth}) we conclude that 
\begin{equation}\label{Eq:power-law}
\left| \langle \hat \tau_I^{x,y}(t)\rangle \right| \sim \frac{1}{\sqrt{{\cal D}(t)}}\approx \frac{1}{{(J_0 t)}^{a}},
\end{equation}
which describes a power-law decay of $x,y$ spin projections, see Fig.~\ref{Fig:ent-growth}(c). We note that the exponent of this power-law is not universal, and, in general is given by $a  =  \xiE s_0$, where $s_0$ is the diagonal entropy~\cite{Polkovnikov} of the initial state per spin~\cite{Serbyn14}. For the state in Eq.~(\ref{Eq:psi-tau}) with all $A_i=B_i=1/\sqrt{2}$, the diagonal entropy attains its maximal value, $\ln 2$ per spin, leading to  an exponent $a=\xiE \ln 2$.

 The dynamics of physical spin operators can be analyzed by expressing them via $\hat \tau_{i}^{\alpha}$, $\alpha=x,y,z$ operators and their products. The terms in that expansion which only involve $\hat \tau^z$ operators will remain unchanged, while any term which involves at least one $\hat \tau_i^{x,y}$ operator, will decay in a power-law fashion. Thus, generic local observables approach their long-time equilibrium values in a power-law fashion. We emphasize that equilibrium values of observables retain the memory of the initial state due to the extensive set of LIOMs. 

Another interesting implication of the dephasing dynamics is that the standard spin-echo protocol can fully recover the state of a given $\tau$-spin~\cite{Bahri:2015aa,Serbyn_14_Deer},  implying that the intrinsic $T_1$ relaxation time remains infinite in the MBL phase. On the other hand, the $T_2$ time induced by the entanglement dynamics with distant spins increases exponentially with the distance to these spins, reflecting the logarithmic dynamics of entanglement growth~\cite{Serbyn_14_Deer}. Of course, in practice one would perform the spin-echo protocol on a physical spin, which only leads to an incomplete recovery of the initial state,  but the revival probability is large at strong enough disorder, when the physical $\hat \sigma_i^z$ operators are close to $\hat \tau_i^z$. 

 Other, closely related experimental signatures of the dephasing dynamics include temporal revivals of local observables~\textcite{Vasseur14}, and double-electron resonance, which can be viewed as a modification of the spin echo protocol, allowing one to probe the dephasing of a given spin by distant spins~\cite{Serbyn_14_Deer}. In addition, power-law decays of various quantities due to the same mechanism have been identified, including mutual information~\cite{Bardarson17}, fluctuations of the out-of-time correlation functions~\cite{HuseOTOC} and fluctuations of the Loschmidt echo~\cite{Serbyn17}.

In the above discussion, we addressed the dynamics of isolated MBL systems. In practice, no system is perfectly isolated from the environment, and therefore it is important to understand how dissipation affects the dynamical behavior. In general, coupling to the bath leads to delocalization and restores slow transport. A classic example is the variable-range hopping: electrons in a solid can hop between localized states, with the mismatch energy being provided by phonon absorption/emission~\cite{MottVRH}. 

Recent experimental studies of MBL were, however, performed in synthetic quantum systems of e.g.\ ultracold atoms or ions, which are free from phonons, since the lattice potential is generated by lasers. Dissipation is still present, with two significant sources being inelastic scattering on lattice lasers, and particle loss for atomic and trapped ion experiments. \textcite{Fischer,Marko16,Levi16}  developed a theoretical approach for describing these kinds of dissipation. They formulated the Lindblad equation in terms of LIOMs, and, having reduced it to a classical rate equation, analyzed the resulting dynamics. It was shown that relaxation of an initial density modulation displays a certain degree of universality, following a stretched-exponential law. In a different direction,~\textcite{Nandkishore14} demonstrated that the spectral function of an MBL system weakly coupled to a heat bath still carries signatures of localization.

\subsection{New numerical and analytical approaches\label{Sec:MBL-methods}}

As we discussed, the theory based on LIOMs provides a natural description of the MBL phase. In the basis of the LIOMs, the eigenstates become product states and the dynamics is reduced to dephasing. Yet, in contrast to the Yang-Baxter-type integrability, where integrals of motion are known exactly, the presence of disorder in the MBL phase precludes an explicit analytic construction of LIOMs. Thus, in order to better understand the properties of LIOMs and the way they become non-local in the vicinity of the MBL transition,  it is necessary to develop new theoretical and numerical tools for constructing highly excited eigenstates.

The low, area-law entanglement of the highly excited MBL eigenstates in 1d allows for their efficient representation by the matrix product state (MPS) ansatz~\cite{SchollwockRMP}. This opens the door to using the density matrix renormalization group (DMRG), originally developed to for the ground states of the one-dimensional systems. (Note, however, that extending DMRG techniques to the excited states is highly non-trivial due to the fact that the level spacing becomes very small at a finite energy density.) Works by \textcite{Pekker15,Sheng15,Pollmann15,Karrasch15,Serbyn16E} provided a proof-of-principle  of the applicability of DMRG algorithms by extracting highly excited eigenstates in MBL system and studying their properties. A more ambitious approach proposed by \textcite{Pollmann15-1,Simon16} uses the matrix product operator ansatz~\cite{SchollwockRMP} to find the unitary $\hat U$, introduced above, which diagonalizes the full Hamiltonian and allows for an explicit construction of LIOMs. 

In a different direction,  the generalizations of the real-space renormalization group to highly excited states provided an alternative set of tools to describe the properties of the MBL phase. We already discussed  in Section~\ref{Sec:MBL-LIOM} the dynamical RG of \textcite{Vosk13} and the real space RG for excited states of \textcite{Pekker14}.  While MPS methods are more suitable for area-law entangled states, the RG-based approaches are also capable of capturing the  structure of the eigenstates with a logarithmic scaling of entanglement.  In particular, works by \cite{Pekker14,Cenke16XYZ} investigated critical properties of the transition between different MBL phases for Ising and XYZ chains, while \cite{VasseurHotChains} studied random SU$(2)_k$ anyon chains. 

The development of new numerical methods  in the two  directions outlined above is an active subject of current research. In addition, tensor-network based approaches can allow access to the long time dynamics of large open systems~\cite{Fischer,Znidaric16}.  Many of the properties of the simplest models of the MBL phase~Eqs.~(\ref{Eq:fermionic})-(\ref{Eq:XXZ}) can be  studied using exact diagonalization. At the same time, the new numerical tools are essential for describing  1d systems with local Hilbert space larger than 2 (bosons, spinful fermions, higher spins) and for studies of phase transitions between  MBL and thermal phases, as well as between different MBL phases.  Finally, the development of tensor-network methods for excited states is necessary for investigating MBL in higher dimensions~\cite{Wahl17}, which is also the subject of current experimental studies, see Section~\ref{Sec:exp}.

\section{MBL-protected phases of matter\label{Sec:MBL-enabled}}

Landau's theory of symmetry breaking gives a description of phases of matter in thermodynamic equilibrium. The fact that MBL systems are not able to reach thermodynamic equilibrium calls for reexamining the notion of a quantum phase of matter in the presence of localization.  Below we discuss the properties of MBL that arise from the presence of additional symmetries. In particular, in the presence of MBL,  discrete Abelian symmetries, such as  a $Z_2$ symmetry, can either be spontaneously broken even in highly excited eigenstates or remain intact, giving rise to distinct MBL phases. In contrast, we will show that non-Abelian symmetries are incompatible with MBL, at least in the sense of having area-law entanglement of eigenstates and a complete set of quasi-local integrals of motion. Next, we will turn to periodically driven (Floquet) systems. We will  discuss that MBL may persist in the presence of periodic driving, resulting in a Floquet-MBL phase. Periodic driving enriches possible kinds of dynamics, and, as a result, MBL can enable Floquet phases of matter with no counterpart in equilibrium systems. Finally, we will briefly discuss the relation between MBL and spin  and Bose glass physics.

\subsection{Symmetries and localization-protected orders  \label{sec:SPT}}

In conventional statistical mechanics, phases of matter can be characterized by the order parameter and its symmetry in thermal (Gibbs) states of the system. The assumption of thermal equilibrium puts constraints on the existence of possible phases and phase transitions. One well-known example is the absence of symmetry breaking at finite temperature in one-dimensional systems with short-range interactions~\cite{Mermin}. The eigenstates of ergodic systems, obeying ETH, are expected to behave similar to the Gibbs states, and to satisfy the same thermodynamic constraints. 

The eigenstates of MBL systems, in contrast, violate ETH, and therefore can exhibit richer order compared to thermal ensembles. Intuitively, as described by \textcite{Huse13L}, localization of excitations that would have destroyed the order in a thermal state, ``protects" the order in individual eigenstates. To illustrate this, and to see how distinct MBL phases may arise, let us consider   a disordered Ising spin chain with $Z_2$ symmetry:
\begin{equation}\label{eq:disordered_Ising}
\hat H_{\rm Ising}=\sum _i J_i \hat\sigma_i^z  \hat\sigma_{i+1}^z +\sum h_i \hat\sigma_i^x + \lambda\sum _i  \hat\sigma_i^x\hat\sigma_{i+1}^x, 
\end{equation}
where $J_i,h_i$ are independent, positive random couplings with non-zero means $\bar{J},\bar{h}$, and variances $\delta J^2, \delta h^2$. The global $Z_2$ symmetry is implemented by the spin-flip operator
$\hat P=\prod_i \hat\sigma_i^x.$

The above model can be mapped to a fermionic model via the Jordan-Wigner transformation; the first two terms map onto a free fermion model with disorder, while the third term corresponds to a four-fermion interaction term. In the ground state, two phases are possible: the ordered phase for $J_i\gg h_i$, which breaks the $Z_2$ symmetry, and the paramagnetic phase for $h_i\gg J_i$, which respects the symmetry. The two phases are separated by the infinite-randomness fixed point described by the strong-disorder real-space renormalization group~\cite{DasguptaMa}. 

Turning to the properties of eigenstates at finite energy density, it is convenient to start from the non-interacting (in the fermionic language) limit $\lambda=0$. In the ordered phase, the excitations are domain walls, which are localized by arbitrarily weak disorder (since the model is non-interacting). Likewise,  spin-flip excitations which are relevant in the paramagnetic phase, are also localized. Upon including sufficiently weak interactions $\lambda\neq 0$, both kinds of excitations may remain MBL~\cite{Huse13L,Pekker14, Kjall14}. Thus, even in the presence of interactions, two distinct phases arise: the ordered phase, characterized by long-range ``spin-glass" order $\langle \hat \sigma_i^z  \hat\sigma_j^z\rangle_\alpha\neq 0$ for $|i-j|\to \infty$ (here $\langle...\rangle_\alpha$ denotes expectation value in an eigenstate $|\alpha\rangle$), and the paramagnetic phase. 

\textcite{Huse13L,Pekker14} argued that the eigenstate transition between the two MBL phases (``spin-glass" and paramagnetic) may be of the same infinite-randomness universality class as the ground state transition. Within the approximation used by~\textcite{Pekker14} this infinite-randomness fixed point remains stable at finite energy density. At the same time, the presence of other delocalization channels which may change the kind of transition (e.g., giving rise to a thermal phase separating the two MBL phases) has not been ruled out. 

The ideas outlined above can be extended to other discrete, Abelian symmetries, and higher-dimensional systems. \textcite{Huse13L, Bauer13} argued that MBL can protect the $Z_2$ topological order in $d=2$ at finite energy density. Further, \textcite{Bahri:2015aa,ChandranKhemani} considered certain models of symmetry-protected topological (SPT) phases, arguing that they can be MBL. The bulk topology can give rise to a protected qubit at the edge, surprisingly, even when the system is very ``hot'' and strongly coupled to the degrees of freedom making up the qubit~\cite{Bahri:2015aa}.

While Abelian symmetries  enrich the variety of possible MBL states as discussed above, the presence of  an unbroken non-Abelian symmetry places strong restrictions on the structure of eigenstates, and is incompatible with  MBL as defined above. More specifically, it is impossible to have the area-law entanglement of excited eigenstates and a complete set of LIOMs ~\cite{PotterSymmetry,Vasseur16,Abanin16}.  For instance~\cite{Abanin16} demonstrated that the area-law entanglement is incompatible with the SU(2) symmetry and the eigenstate entanglement entropy must scale at least logarithmically with system size. The interplay of MBL and different discrete non-Abelian symmetries was also addressed in Refs.~\cite{Vasseur16,Vasseur17,Fidkowski17}.  Searching for possible non-ergodic phases beyond MBL in the presence of non-Abelian symmetries is a promising research direction. 

\subsection{Many-body localization in Floquet systems}\label{subsec:MBL-Floquet}

We proceed by discussing another application of MBL:  in periodically driven (Floquet) systems,  MBL can prevent heating to  an infinite temperature state, opening up the possibility of having new non-equilibrium Floquet-MBL phases. 

Subjecting a physical system to an external, time-periodic perturbation, e.g., with lasers, is a powerful experimental tool. Recently, this tool has been used to control and engineer properties of synthethic quantum systems, leading, e.g., to the realization of topological Bloch bands in systems of ultracold atoms~\cite{Cooper2018TopologicalBands}. Furthermore, it was shown theoretically that single-particle Floquet systems exhibit a rich variety of novel topological states, which are not possible in equilibrium~\cite{Kitagawa10}. A prominent example of such topological, ``Floquet-only" phases is a 2d system with protected edge states, but no bulk bands with non-zero Chern numbers~\cite{Rudner13}. 

Real systems are interacting, and intuitively one expects that ``shaking" an interacting system would almost inevitably cause heating. Such heating arising from energy absorption in interacting systems will wash out interesting topological features, such as edge states, thus being a central obstacle in the field of Floquet engineering. This intuition was recently made precise in Refs.~\cite{Ponte14,Alessio14,Lazarides14}, which argued that ergodic systems satisfying ETH heat up indefinitely under driving. 

However, MBL systems break ETH and therefore may avoid heating~\cite{Ponte14}. Refs.~\cite{Ponte15,Lazarides15,Abanin20161} established that MBL can indeed remain stable in periodically driven systems, as long as the frequency of the drive is sufficiently high. At low driving frequency, in contrast, delocalization is inevitable, even for a drive with a small amplitude~\cite{Abanin20161}. The {\it Floquet-MBL phase} is characterized by a complete set of LIOMs, area-law entanglement of Floquet eigenstates, but most importantly, MBL prevents heating to an infinite temperature, opening the door to stabilizing Floquet-only phases in isolated systems.  Signatures of the Floquet-MBL phase have been observed in a recent experiment with ultracold atoms~\cite{Bordia17}. 

One example of an MBL-enabled  Floquet-only phase is the discrete time crystal~\cite{Khemani16,Else16}, characterized by the breaking of discrete time-translation symmetry of the drive ($t\to t+nT$). As a result, at long times local observables generally do not relax, instead exhibiting persistent oscillations at multiple integer of the driving period. First experimental signatures have been reported in NV-centers spin system in diamond and in trapped ionic systems~\cite{Choi16DTC,Zhang16DTC}. However, these systems exhibit long-range interaction, which precludes localization,  see discussion in Section~\ref{Sec:exp:ions}. \textcite{Ho17} demonstrated that the  observed transient time-crystalline signatures stemmed from the parametrically slow relaxation caused by rare two-spin resonances (``critical time crystal"). 

Another, qualitatively different example of an interacting, Floquet-only phase is the anomalous Floquet insulator. This is  a two-dimensional phase of matter which is MBL in the bulk, yet has topologically protected chiral edge states. The stability of this phase was shown in~\cite{Nathan17}, while ~\cite{Gross12,Po16} discussed the topological invariants which protect this and related phases. Finally, we note that Refs.~\cite{Else16-1,Curt16-1,Curt16-2,Roy17,Po16,Morimoto16} put forward (partial) classifications of distinct Floquet-MBL phases in the presence of additional symmetries. Theoretical and experimental investigations of new Floquet-MBL phases, their physical properties, and topological invariants remain a subject of active research.

\subsection{Many-body localization, spin and Bose glasses}

There are apparent similarities between the physics of glasses, which has been a subject of intense study for many years, and the phenomenon of many-body localization. In particular, both phenomena involve breaking of ergodicity. The goal of this Section is to compare the two phenomena and clarify the essential differences between them. We discuss two kinds of glasses, which can occur in quantum systems: a spin glass, and a zero temperature phase commonly referred to as a Bose glass. 

\subsubsection*{1. Spin glass \label{sec:SpinGlass}}
A spin glass is a low-temperature phase found in certain disordered (either classical or quantum) spin models at sufficiently low temperatures. In the ideal case, there is a thermodynamic phase transition at a critical temperature below which the system breaks ergodicity. As discussed below, this ergodicity breaking has a different character compared to the case of MBL, since a glassy system can remain ergodic within separate parts of the phase space. More generally (even if there is no sharp transition), spin glasses are characterized by a wealth of unusual dynamical phenomena, including very slow dynamics of observables (e.g., magnetization) and memory (aging) effects in a  quench experiment~\cite{SpinGlassReview86}. While MBL is also associated with slow dynamics and breaking of ergodicity, the origin of these effects is very different, allowing to make a sharp distinction between MBL systems and spin glasses.

 The source of the unconventional properties of glasses is frustration.  The fact that the interaction terms cannot all be minimized simultaneously  results in a large   number of low-energy states that are separated  by  energy barriers that increase with the system size; this leads to the characteristic ``rugged'' energy landscape in phase space. When coupled to a bath maintained at sufficiently low temperature, the large barriers prevent the system from exploring the entire phase space. The broad distribution of  energy barrier heights leads to a broad distribution of relaxation timescales in glasses. Note that this basic fact about glasses is true regardless of whether the microscopic degrees of freedom are classical or quantum (e.g. quantum spins). As long as there is a broad distribution of energy barriers, fluctuations induced by coupling to a (cold) thermal bath would generate glassy dynamics. Hence, classical glasses are not only robust with respect to coupling to an external bath, but their dynamics is in fact often generated by such coupling.

In stark contrast, MBL is a fully quantum phenomenon, which  does not require frustration, and instead relies on  the discreteness of the spectrum, or equivalently, on interference effects. The intuitive criterion for MBL is that the transition rate between two many-body configurations that are very close in energy is much smaller than the many-body level spacing, resulting in the absence of resonances between such configurations. Coupling to an external bath destroys the interference effects that ensure the discrete spectrum and therefore generally also destroys MBL, see  Ref.~\cite{Fischer} and Section~\ref{Sec:open}. We can therefore make a distinction between an MBL system and a glass by studying how its respective dynamics is affected by coupling to a low temperature bath. MBL relies on the system being isolated from the environment, whereas glassiness does not.  Of course,  the presence of frustration may help MBL, but it is not essential for its existence.

The above discussion implies another important difference between the two phases. While certain observables fail to relax (or relax very slowly) in a glassy system, the information retained in these observables is completely classical. In an MBL system, on the other hand, it is possible to recover local quantum information (i.e. local phase information of a q-bit) using spin echoes after arbitrary long times~\cite{Bahri:2015aa,Serbyn_14_Deer}.

While the basic mechanisms of an MBL and a spin glass phase are qualitatively different, there is an interesting, and largely unexplored question regarding their possible coexistence.  More generally, it is desirable to develop an understanding of dynamical properties of isolated quantum systems with glassy classical energy landscape subject to unitary quantum dynamics.

To this end, one may consider the spin-$1/2$ Edwards-Anderson model with a transverse field, which gives rise to quantum dynamics:
\begin{equation}\label{eq:EA_quantum}
\hat H_{\rm EA-q}=-\sum_{ij} J_{ij} \hat \sigma_i^z \hat \sigma_j^z+\Gamma\sum_i \hat \sigma_i^x,
\end{equation}
where $J_{ij}$ are random couplings.  The statistics and dependence of interaction $J_{ij}$ on the distance between spins  determine  the phase diagram of the  corresponding classical  model. For instance, infinite-ranged interactions where the distribution of $J_{ij}$ is independent of $i,j$ lead to the Sherrington-Kirkpatrick model~\cite{SKmodel}. On the other hand, embedding spins into a $d$-dimensional lattice with only nearest-neighbor $J_{ij}\neq 0$ would describe a short-range spin glass. While the classical~($\Gamma=0$) phase diagram of model in Eq.~(\ref{eq:EA_quantum}) is known in many cases, much less is known about the quantum model. What is the nature of quantum eigenstates? Can this model be in the MBL phase? 

\onlinecite{PalSpinGlass} investigated these issues for mean-field spin glass models with $p$-spin interactions (the Sherrington-Kirkpatrick model corresponds to the case $p=2$). They found a parameter regime in which eigenstates at a given energy density cluster into groups with different values of observables, such that ETH is satisfied within one cluster. While an MBL phase is impossible in such systems due to infinite range of interactions, the behavior of the model in Eq.~(\ref{eq:EA_quantum}) with short-range interactions may allow for an MBL phase or non-ergodic phase similar to one found by~\cite{PalSpinGlass}. Investigating these issues comprises a promising direction for future research.

\subsubsection*{2. Bose glass \label{sec:BoseGlass}}
Bose glass (BG) is the term used to describe insulating quantum phases of interacting bosons in a disorder potential at {\em zero temperature}.
Such phases are found in a broad range of physical systems, including $^4$He in porous media, cold atoms in disordered optical lattices~\cite{BoseGlassColdAtoms07,Meldgin}, thin superconducting films~\cite{Goldman89,Hebard90}, and disordered magnets~\cite{BoseGlassMagnet12}. Theoretically, the existence of the Bose glass was established in a pioneering work of~\cite{Giamarchi88}, who used perturbative renormalization group to analyze the transition between the BG and fluid phase in one dimension. The BG in higher-dimensional systems was studied by~\cite{Fisher89}, who derived the critical exponents for the BG-superfluid transition. 

Because Bose glasses are zero-temperature ground states, while MBL is a property of highly excited states, these two phenomena refer to different parts of the spectrum and are not directly related. However, it is natural to ask if having a Bose glass ground state necessarily implies many-body localization  at low non-vanishing temperatures. This question was investigated for disordered weakly interacting bosons in one and two dimensions by~\textcite{BoseMBL1,BoseMBL3},  who argued that the BG phase is smoothly connected to an MBL phase as the temperature is increased. \onlinecite{BoseMBL1,BoseMBL2} analyzed the critical temperature for the MBL-delocalization transition (equivalently, the position of the many-body mobility edge) as a function of the disorder strength and interactions. Furthermore, \onlinecite{BoseMBL2} studied MBL for strongly interacting bosons in 1d, also finding that the zero-temperature BG phase at both weak and strong disorder is smoothly connected to MBL phase at finite temperature. 

It would be interesting to confront these theoretical arguments for the smooth connection between BG and MBL phases with an experimental test.  In addition, the connection between the BG and MBL phases when interactions are strong, is an interesting open question that remains unexplored.

\section{Delocalization transition \label{Sec:Tran}}

The breakdown of many-body localization upon changing the disorder strength or some other control parameter provides an intriguing opportunity to study the emergence of thermalization in a quantum system, possibly with the control afforded by proximity to a critical point.~\footnote{We assume the direct transition between MBL and thermal phases throughout this Section and do not consider possibility of e.g.\ intervening glassy phase at finite temperature.}
 In the MBL phase, quantum information encoded in local observables is protected and affects the dynamics at arbitrarily long times. On the other hand in a thermalizing system quantum information is lost to non-local degrees of freedom and the remaining slow modes are described by classical hydrodynamics. Thus investigating the transition can shed light on the elusive boundary between quantum and classical behavior in interacting systems.

In addition to  providing insights into  the mechanism of quantum thermalization, the  delocalization transition represents a new type of quantum phase transition, which differs in crucial aspects from both thermal  and zero-temperature quantum phase transitions. One important distinction is that  unlike in conventional phase transitions, there need not be any thermodynamic signatures of the delocalization transition, and it can be manifested only in dynamical quantities, like energy conduction or entanglement propagation. 
 Another unique feature is that the MBL transition must involve a dramatic change in the real space entanglement structure of many-body eigenstates. At the critical point, the entanglement entropy of energy eigenstates  changes from area law on the localized side to volume-law entanglement entropy, consistent with the thermodynamic entropy, on the thermal side of it.  In contrast, the usual quantum critical points mark the  transition between two ground states with area-law entanglement. In spite of those differences,  important insights into the MBL transition have been gained by adapting renormalization group ideas.

\subsection{ Renormalization group approach to the MBL transition \label{sec:RG}}

The MBL phase is stable against thermalization due to the extreme rarity of resonances when the disorder is sufficiently strong. The prevailing view of the delocalization transition is that it is driven by the emergence of resonant clusters. As the strength of randomness is reduced, resonant clusters start to occur more frequently in the system. At the critical disorder strength a critical cluster grows to encompass the entire system. The renormalization group (RG) approaches to the MBL transition attempt to describe the fluctuations that give rise to such a critical cluster at multiple scales and predict how they impact  properties of the system near the critical point. 

Different RG schemes were proposed to describe this process~\cite{AltmanRG14,Potter15X,Dumitrescu17,MullerRG,huse,GVS}. The RG scheme in  \textcite{AltmanRG14} starts from a phenomenological coarse grained description of a one dimensional system as a chain of regions with varying local character: some regions behave locally as insulators and others have a local thermal character. The RG scheme then attempts to describe the competition between growth of the thermal regions as they hybridize with nearby clusters on the one hand, and their potential isolation by surrounding MBL regions on the other hand. The RG scheme of \textcite{Potter15X} takes a more microscopic starting point, operating on spin-$1/2$-like degrees of freedom. The scheme attempts to construct the critical cluster from the bottom up: identifying resonant pairs of spins, joining them into resonant mini clusters, then joining those to even larger clusters and so on. In the MBL phase this process stops with a finite cluster, when no more resonant clusters can be merged, while in the delocalized phase a resonant cluster eventually encompasses the entire system.

In both cases, the key scaling variable that identifies the degree to which a cluster is localized or thermalizing is a ratio  $g=\Gamma/\Delta$ between the "decay rate" $\Gamma$ associated with the rate of information loss across the cluster and the many-body level spacing. If $g\gg 1$ then the cluster is resonant and we say it is locally thermalized, while if $g\ll 1$ then the cluster is said to be localized. 
This ratio is reminiscent of the Thouless conductance, defined in non-interacting systems by the ratio between a single particle relaxation rate (the Thouless energy) to the level spacing, which is the central object of the scaling theory of  Anderson localization~\cite{ScalingTheory}. The ratio $g$ can  be viewed as a ``many-body Thouless parameter''. A microscopic version of the many-body Thouless parameter was introduced in \cite{Serbyn15} and used to diagnose  the MBL transition and its properties.

An important difference from the single particle case is that the many-body level spacing is  exponentially small with the cluster size $l$. For a spin-$1/2$ chain $\Delta \sim e^{- (l/a)\,\ln 2  }$ near the middle of the spectrum (infinite temperature) and more generally $\Delta\sim e^{-s\,l/a}$, where $s$ is the entropy density and $a$ the lattice spacing. In the MBL phase the relaxation rate of a cluster is also exponentially small with the cluster size (distance to conducting leads), $\Gamma\sim e^{-l/\ell_*}$, where $\ell_*$ is a localization length. Comparison between the two exponentially small scales imposes a stringent cutoff for the localization length associated with propagation of information through an MBL system. For the system to remain localized with $g\ll 1$ we must have $\ell_*< a/s$.

From the existence of a finite (non critical) localization length $\ell_*$, it is tempting  to infer that the MBL transition must be first order. However, the RG schemes \textcite{AltmanRG14,Potter15X} do find a critical fixed point with a diverging length scale $\xiRG\sim |g-g_c|^{-\nu}$ with $\nu\approx 3.3$. Nonetheless the transition appears as a non-critical jump when viewed through the lens of typical observables, measured on a single sample.  It was pointed out in \textcite{AltmanRG14} that to observe the growing critical cluster on the MBL side of the transition one must measure  average values. As we shall see in the next section, the critical behavior is much more accessible to experiments done on the thermal side of the MBL phase transition.

It is interesting, and encouraging, that in spite of the different philosophies underlying the
two RG schemes, they give similar predictions for
the critical properties, including the correlation length
exponent, dynamical exponents and more. The critical exponent $\nu\approx3.3$ found by both RG schemes is consistent with the Harris bound $\nu>2/d$ \cite{Harris,Chayes}.~(\textcite{Chandran2015} argued that Harris bound holds  for the MBL critical point  in spite of its unconventional aspects). By contrast, exact diagonalization results for small spin chains give the appearance of scaling with a critical exponent that violates the bound,  $\nu\approx 1<2/d$ \cite{Kjall14,Alet14}. 
The violation of the Harris criterion by the exponent extracted numerically for XXZ spin chains may be due to the fact that systems that can be diagonalized exactly are too short to be in the asymptotic scaling regime.

\begin{figure}[b]
\begin{center}
\includegraphics[width=1.0\linewidth]{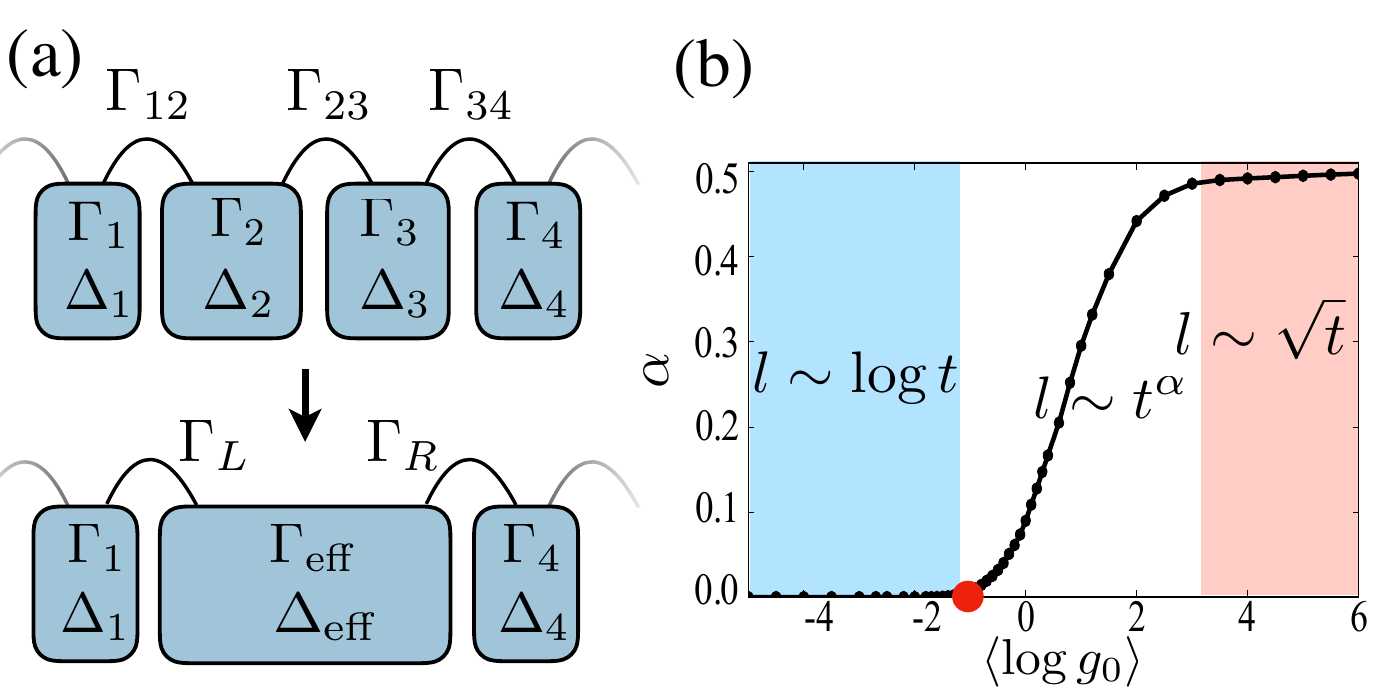}
\caption{\label{Fig:RG} (a) RG description starts with a collection of blocks each characterized by the hybridization rate and level spacing. Within one RG step two blocks with the strongest coupling rate are merged together resulting in a block with new effective parameters.
(b) The inverse dynamical exponent $\alpha$  vanishes in the MBL phase and interpolates continuously between zero and $1/2$ in the thermal phase. 
}
\end{center}
\end{figure}

At this point it is worth noting that the RG scheme has identified two distinct localization lengths  $\xiRG$ and $\ells$, where the former length diverges and the latter stays finite at the transition. The divergent localization length $\xiRG$ is rather hard to observe inside the insulating phase as it is not manifested in typical values (at least of the quantities considered in Ref.~\cite{AltmanRG14}). Instead one must measure average values which are sensitive to rare events and therefore may require very large sample sizes to converge. While the relation between the phenomenological lengthscales $\xiRG$, $\ells$ and the microscopic parameters $\xi,\xiJ, \xiE$ is unclear at the moment, we note that (i) $\xiE$ controls the entanglement spreading, as does $\ells$ and therefore they are likely proportional to each other; and (ii) the localization length $\xi$ (or its average) may diverge at the transition, just like $\xiRG$. 

The entanglement structure of eigenstates is intriguing and helps in developing a theoretical picture of the MBL transition, however it is not accessible to experimental measurement. In the next subsection we discuss critical relaxation dynamics and transport properties which can serve as realistic probes of the critical point in experiments.

\subsection{Subdiffusion and Griffith regions}

The renormalization group approaches \cite{AltmanRG14,Potter15X} give  explicit predictions for the  transport properties near MBL transition. The scheme of \textcite{AltmanRG14} allows to plot the energy relaxation time $\tau_{\text{tr}}=(l/l_0)\Gamma^{-1}$ of clusters versus their average size $l$ to obtain the dynamical exponent  $1/\alpha$, $l\sim \tau_\text{tr}^\alpha$. In the  MBL phase one finds a logarithmic relation $l\sim \ells \ln(\tau_{\rm tr})$, hence $\alpha=0$ in Fig.~\ref{Fig:RG}(b). While naively one would expect to see the diffusive scaling throughout the thermal phase, $l\sim \sqrt{D\tau_{\text{tr}}}$,  with a diffusion constant that vanishes at the critical point, the RG scheme observes $\alpha=1/2$ only far from transition. Closer to the transition, however, both RG schemes \cite{AltmanRG14,Potter15X} as well as earlier numerical studies \cite{Reichman15,Demler14,Znidaric16}, found subdiffusive transport $\tau_{\text{tr}}\sim l^{1/\alpha}$ with the inverse dynamical exponent  $\alpha$ varying continuously. The {dynamical} exponent $z\equiv 1/\alpha$ diverges at the critical point as $z \sim\xiRG \sim (g_0-g_{0c})^{-\nu}$, with $\nu\approx 3.3$, while at criticality the transport shows exponential scaling just as in the insulating phase. 

The sub-diffusive scaling is understood to be a result of rare critical inclusions in the thermal phase. Singularities due to rare regions have been first discussed by  \textcite{Griffiths69,McCoy69} in the context of conventional phase transitions of random spin systems and since then are known as Griffiths effects. The key to the contribution of rare regions to the transport near the unconventional MBL critical point is the balance between the  low probability of finding such a long region and the  long delay it would affect as a bottleneck to transport.
The probability to find a long critical cluster of length $l$ falls off exponentially with its length. For a system of length $L$ near the critical point it is $P_L(l)\sim (L/\xiRG)\exp(-l/\xiRG)$, where $\xiRG$ is the  correlation length that diverges at the critical point. 
Thus, the longest critical inclusion we are likely to find in a system of length $L$ (with probability of order 1) is $l_m(L)\sim \xiRG\ln (L/\xiRG)$. If such a rare inclusion serves as the dominant bottleneck on transport, it leads to a relaxation time  $\tau(L)\sim \exp[l_m(L)/\ells]\sim L^{\xiRG/\ells}$. From this scaling we can read off the dynamical exponent $1/\alpha \sim \xiRG/\ells$ whose singular behavior at the MBL transition directly reflects the divergence of the correlation length. The RG results shown in Fig.~\ref{Fig:RG}(b) agree with the conclusion of these elementary considerations, showing that $\alpha \sim (g_0-g_{0c})^{\nu}$. 

The dynamical scaling affects not only the transport through the system, but also relaxation of global observables, such as the decay of a charge density wave imposed on the system \cite{Bloch15}. In this case the rare regions lead to power-law relaxation instead of the exponential decay that would otherwise be expected from a non-conserved operator.

So far we have discussed only the effects of rare regions on transport on the thermal side of the critical point. The implication of such regions to the AC conductivity in the MBL phase have been investigated in Ref.~\cite{Gopa-15}.  However, those effects compete with other types of many-body resonances and it is not clear what the relative contribution of the different effects is. Hence relaxational dynamics in the MBL side of the transition is not well understood at this point.

Our discussion of the MBL critical point so far pertained to one dimensional systems. How do these ideas generalize to higher dimensions? Na\"ively, the thermalization criterion based on the many-body Thouless parameter defined above  always predicts thermalization in a sufficiently large system. Indeed, if the thermalization rate $\Gamma$ falls off exponentially  with the linear size of the system $L$, the level spacing decreases as $\exp(-s L^2)$ and is therefore always much smaller as $L\to \infty$. \textcite{Roeck} have turned this insight into a more systematic bootstrap scheme, which suggests that the MBL phase may not be stable to the presence of a sufficiently large but finite thermal inclusion.  On the other hand, these arguments rely on the assumption that a metallic region is able to ``perfectly thermalize'' a nearby insulator, hence the issue of MBL (in)stability in higher dimensions is far from being settled. In Section~\ref{Sec:exp} we return to the discussion of experiments which display signatures of MBL in two-dimensional lattice models.    

\section{Experimental developments \label{Sec:exp}}

Probing for MBL is challenging in experiments, as the system under investigation has to be isolated from any thermal environment. This makes it very hard to observe MBL in standard materials, as almost all of them are connected to thermal reservoir during cooling and experimental cycles. Quantum simulators based on ultracold atoms and ions have therefore been among the first systems in which MBL could be observed, owing to their almost perfect isolation and small couplings to the outside world. There is currently also a very active search for MBL in real materials. It requires finding degrees of freedom which are extremely weakly coupled to the standard thermal bath of phonons present in a solid. NV centers of spins in diamond crystals and electrons in disordered superconducting films have shown interesting possible signatures of localization, but research is still ongoing to unravel the complex interplay of phenomena in these systems. One should note that in any experiment, even the ones based on almost ideally isolated systems of ultracold atoms or ions, a small coupling to the environment is inevitable. The question of how such couplings can affect MBL and the associated phase transition will therefore be discussed in a separate subsection. In addition, finite size effects of smaller experimental samples can play an important role in the interpretation of the data.

\subsection{MBL with ultracold atoms}
When exploring MBL, we recall that one is trying to identify a new phase and phase transition at high energy densities, far away from the ground state of an interacting many-body system.   Earlier related experiments on the ground state physics of disordered interacting particles had  been  successful in establishing evidence for the existence of a Bose glass phase  (see Section \ref{sec:BoseGlass} and Refs.~\cite{BoseGlassColdAtoms07,Meldgin}) and disorder induced localization of a metallic state~\cite{DeMarco15}. To study MBL and the associated transition, experiments have mostly resorted to preparing highly out-of-equilibrium initial states and probing for their subsequent time evolution. As discussed in Section~\ref{Sec:MBL-dyn}, an MBL system will evolve into a stationary state, in which some local observables will assume non-thermal expectation values, whereas an ergodic thermalizing system would exhibit thermal expectation values for all local observables. The presence of such non-thermal local observables directly indicates a non-ergodic evolution of the system and therefore can be used as a way to identify the localized non-thermal phase. Note that it is very hard to show that a system is thermalized, as this would require demonstrating that all local observables are thermal. The opposite, demonstrating localization, can in contrast be rather straightforward: a {\it single} local observable with a non-thermal value is sufficient to show this.

\begin{figure}[t]
\begin{center}
\includegraphics[width=\columnwidth]{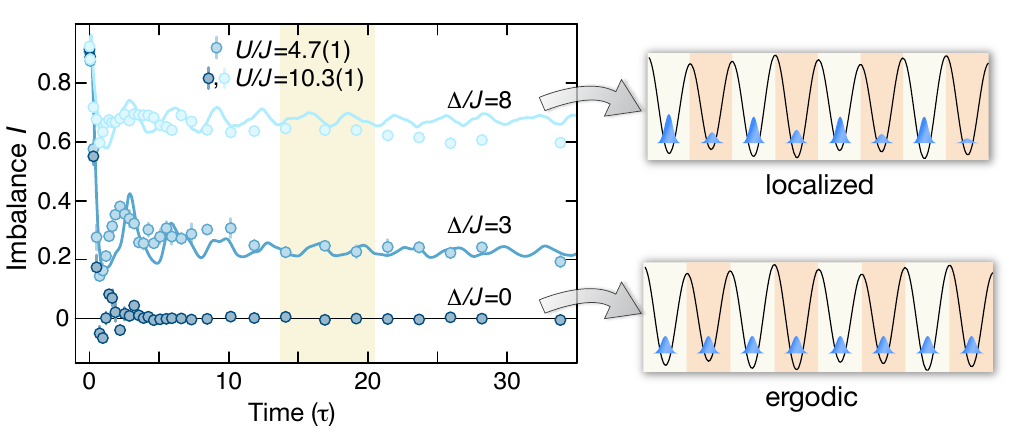}\\
\caption{ \label{Fig:ImbalanceEvolution}
Non-thermalizing out-of-equilibrium evolution of an initial density wave in the presence of a quasiperiodic detuning potential in the interacting Aubry- Andr\'e  model (see Eq.~\ref{Eq:InteractingAA}). Time traces of the imbalance $I$ for various strengths of the detuning potential $\Delta$. Points are experimental measurements, averaged over six different phases $\phi$ of the quasiperiodic detuning lattice. Lines denote DMRG simulations that take into account the trapping potential and the averaging over neighboring tubes, which are present in the experiment \cite{Bloch15}. 
}
\end{center}
\end{figure}

Initial experiments on MBL with ultracold atoms at high energy densities were carried out using one-dimensional Fermi-Hubbard chains of interacting spin-mixtures of two (hyperfine) spin components \cite{Bloch15}.  In order to realize a detuning landscape for the atoms, a quasiperiodic potential was applied, giving rise to the following Hamiltonian in 1d:
\begin{multline}
\hat H = -J \sum_{i,\sigma} \left ( \hat c_{i,\sigma}^\dagger \hat c_{i+1,\sigma} + \text{h.c.} \right ) + \\ \Delta \sum_{i,\sigma} \cos \left ( 2\pi \beta i + \phi \right ) \hat n_{i,\sigma} + U \sum_{i} \hat n_{i,\uparrow} \hat n_{i,\downarrow}. \label{Eq:InteractingAA}
\end{multline}
Here $\hat c_{i,\sigma} (\hat c^\dagger_{i,\sigma})$ denote the fermionic creation (annihilation) operators for a particle in spin state $\sigma=\uparrow,\downarrow$ on lattice site $i$, $U$ is the  onsite interaction strength and $\Delta$ denotes the strength of the quasiperiodic detuning potential. A Feshbach resonance between the atomic spin states allows one to tune the interaction between the particles, $U$, enabling one to directly compare the evolution of repulsively-, attractively- and non-interacting systems starting from the {\it same initial state}. 

For the non-interacting case, $U=0$, the model realizes the celebrated Aubry-Andr\'e (AA) transition, exhibiting Anderson localization for $\Delta/J>2$, which can serve as a well understood reference point for the problem. Furthermore in 1d and for infinitely strong interactions $U\rightarrow \infty$, the system also maps back onto the non-interacting AA problem, if initially no doubly occupied sites are present in the system.  This work, together with theoretical results~\cite{Iyer13,Michal14}, indicates that MBL can also occur in system with quasiperiodic detuning potentials. The question of how the transition in this case is different from the generic MBL transition is a subject of current research~(see e.g.~\cite{Lev17,Pix17,Khe17,li2015,xiaoli2017}).

In experiment~\cite{Bloch15}, the system is initially prepared in a density-wave state, with particles predominantly occupying even sites. The subsequent time evolution of the state is monitored (see Fig.~\ref{Fig:ImbalanceEvolution}), keeping track of the remnant density wave. This is quantified through the imbalance $I=\langle (N_e-N_o)/(N_e+N_o) \rangle$ ($N_e$ and $N_o$ being the even an odd site populations of the system), in analogy to the visibility of an interference pattern in optics. For weak quasiperiodic detuning potential, the imbalance relaxes rapidly -- compatible with a thermalized state of the system. However, above a critical detuning strength, larger than the one in the non-interacting system, the imbalance saturates to a non-vanishing value (see Fig.~\ref{Fig:ImbalanceEvolution}). This is incompatible with thermalization and indicates a localized phase, since a thermal phase occupies even and odd sites with equal probability. Subsequently, the delocalizing effect of coupling many 1d systems subject to identical quasi-periodic potential was experimentally studied in Ref.~\cite{Bloch16}.  In a different direction, \textcite{Lukin18} recently observed logarithmic spreading of entanglement in a small Bose-Hubbard chains subject to a quasiperiodic potential, consistent with the theoretical picture described above.

\begin{figure}[t]
\begin{center}
\includegraphics[width=0.95\columnwidth]{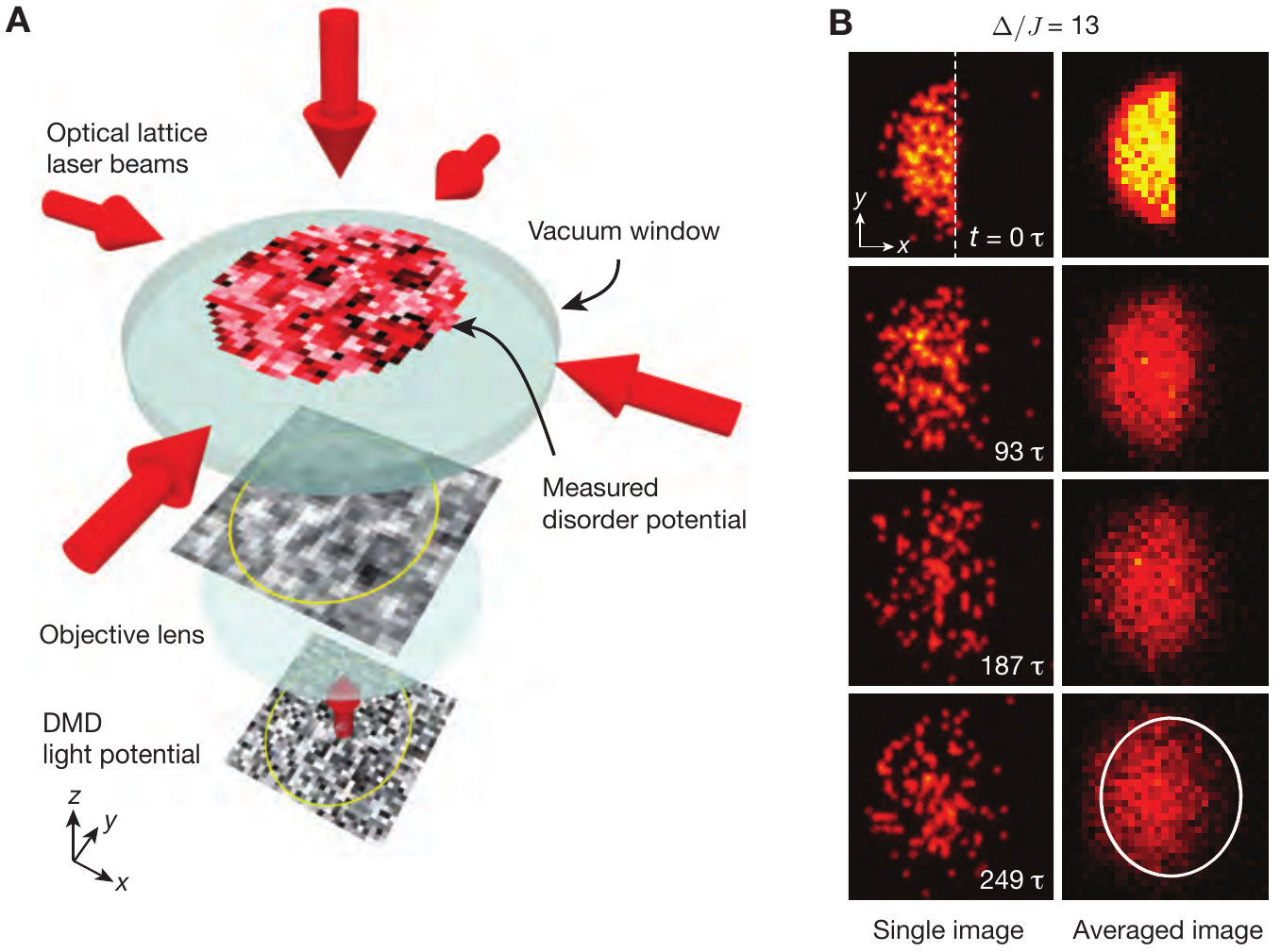}\\
\caption{ \label{Fig:DomainWall2D}
Probing many-body localization in two dimensions. {\bf (A)} Almost arbitrary disorder potentials of light are projected onto an ultracold bosonic atom cloud. The subsequent quantum evolution of an initial non-equilibrium state can then be tracked in the experiment. {\bf (B)} In the experiment an initial domain wall of a bosonic Mott insulator is prepared ("half circle" in images). Even for long evolution times of $\simeq 250$ tunneling times, the system fails to thermalize, indicated by the remnant domain wall still visible in the experiment. In contrast, a thermalized state would not carry any
information about the initial state of the system \cite{Bloch16-2}.
  }
\end{center}
\end{figure}

However, first experiments in a system of two-dimensional interacting bosons exposed to a {\em two-dimensional disorder pattern} seem to also indicate the presence of a localized phase that is reached above a critical disorder strength \cite{Bloch16-2}. Here again the non-equilibrium dynamics of the system was used to probe for a non-thermal evolution, by monitoring the time dynamics of a domain wall in the density of the system (see Fig.~\ref{Fig:DomainWall2D}).

\subsection{MBL with ultracold ions \label{Sec:exp:ions}}

Experiments using one-dimensional strings of ten ultracold ions were used to implement the disordered transverse field Ising model with long-range interactions:

\begin{equation}\label{eq:H_trapped_ions}
 H_{\text{Ising}} = \sum_{i<j} J_{i,j}  \hat\sigma_i^x \hat\sigma_j^x + \sum_i h_i \hat \sigma_i^z + B \sum_i \hat \sigma_i^z. 
\end{equation} 

A specialty of the experment were the long-range interactions between spins, which decay algebraically with distance $J_{i,j}= J_{\text{max}}/|i-j|^\alpha$ and exhibit a tunable decay exponent $\alpha=0.85\ldots1.81$. Random onsite disorder $h_i$ was generated by spin dependent AC Stark shifts of a laser beam and sampled from a uniform distribution $h_i  \in [ -\Delta, \Delta]$. Starting from an initial N\'eel state, the system was evolved in time and exhibited a stationary magnetization above a critical disorder strength, evidencing the presence of a localized phase \cite{Monroe16}. 

An interesting additional feature of the experiment was the measurement of entanglement in the system through the quantum Fisher information~\cite{QFI}. Whereas the non-disordered system showed an initial rapid increase of entanglement, but no subsequent  growth,  the quantum Fisher information of the interacting system exhibited an increase even for intermediate evolution times. Such an intermediate time increase of the quantum Fisher information is consistent with the logarithmic growth of entanglement entropy due to the dephasing between LIOMs (see Section \ref{Sec:MBL-dyn}), whereas a simple non-interacting Anderson insulator does not exhibit such a continued growth of entanglement as a function of time.

Theoretically, sufficiently long-range hopping and interactions are known to destroy single-particle~\cite{Anderson58,Levitov90} and many-body localization~\cite{Burin06,Yao14Dipolar,Burin15,Gutman16}. In the single-particle problem in $d$ dimensions with on-site disorder and power-law hopping $t_{ij}\propto {1}/{r_{ij}^\alpha}$, localization is destroyed for $\alpha\leq d$ due to inevitable resonant transitions, which occur between remote sites in the limit of strong on-site disorder~\cite{Anderson58,Levitov90}. \onlinecite{Burin06} considered a many-body problem with on-site disorder and power-law hopping and interactions decaying with the same exponent $\alpha$, and showed that hierarchical resonances are inevitable and prohibit many-body localization provided $\alpha<2d$. Further, ~\cite{Yao14Dipolar,Burin15,Gutman16} investigated fermionic and spin models in which hopping (or, equivalently, flip-flop processes in spin models) and interactions decay as power laws with different exponents $\alpha,\beta$, and identified the regimes in which hierarchical resonances destroy MBL. 

Interestingly, \cite{Burin15} showed that the model of Eq.~(\ref{eq:H_trapped_ions}) realized in the trapped ion experiments delocalizes for $\alpha<{3d}/{2}$, which is in an apparent contradiction with the experimental observation of localization  in the range  $0.85<\alpha<1.81$. We note, however, that the hierarchical resonances occur at large lengthscales, exceeding the size of the realized ion chains. Thus, the fact that delocalization was not observed at $\alpha<3/2$ in the experiment likely stems from the pronounced finite-size effects. Indeed, numerical simulations of the experiment~\cite{WuSarma} confirmed the absence of relaxation in small ion chains.

We note that \cite{NandkishoreSondhi} recently pointed out an intriguing possibility that MBL can persist at low temperatures in systems where long-range interactions induce charge confinement (e.g., in one-dimensional systems where interactions grow proportional to the distance between particles). 

\subsection{MBL with superconducting circuits}
\begin{figure}[t]
\begin{center}
\includegraphics[width=0.65\columnwidth]{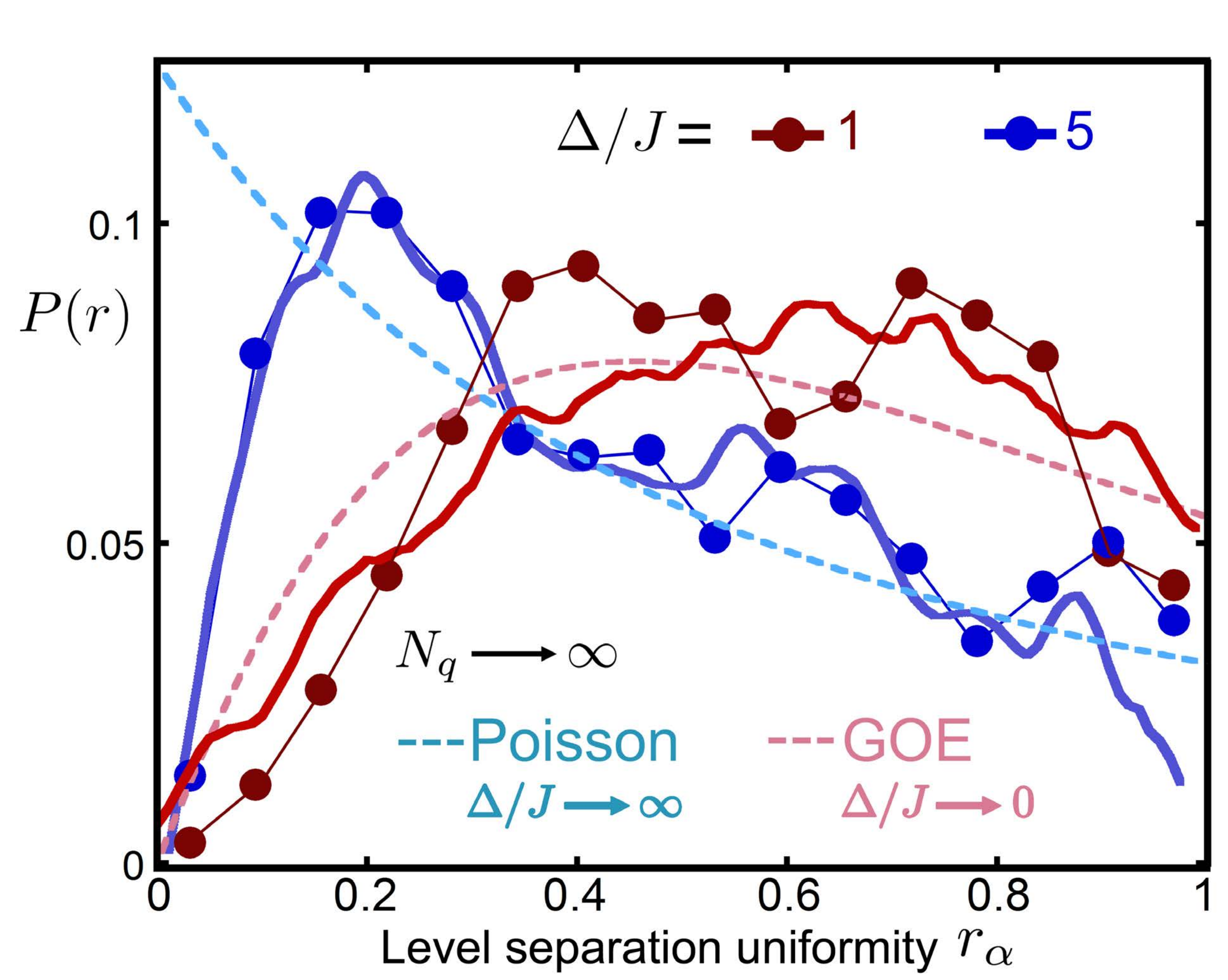}\\
\caption{ \label{Fig:SC-LevelStatistics}
Level spacing statistics obtained in a system of two interacting photons on nine lattice sites and a local disorder potential using many-body spectroscopy. For weak disorder the system exhibits a distribution resembling the one of a Wigner-Dyson Gaussian orthogonal ensemble, whereas for stronger disorder strengths a change towards the one of a Poissonian distribution is observed. Deviations at very small level spacings $r$ are attributed to the finite size and small number of particles in the system \cite{roushan2017}.
  }
\end{center}
\end{figure}

Superconducting qubits have emerged as another powerful platform for tunable and isolated quantum-many body systems~\cite{Houck}. Recently localization signatures for two interacting photons on a nine-site large lattice modeled by a disordered Bose-Hubbard model were obtained through a novel many-body spectroscopy technique. This technique has enabled one to retrieve the many-body eigenenergies of the system and thereby obtain information on the level statistics of the underlying Hamiltonian \cite{roushan2017}. Already at the level of two particles it was found that the level statistics parameter characterizing the average ratio of adjacent level spacings has markedly different properties in the disordered vs the non-disordered case. Whereas the non-disordered system exhibited a distribution compatible with the Wigner-Dyson Gaussian orthogonal ensemble, for increasing disorder a more Poissonian shaped distribution was found (see Fig.~\ref{Fig:SC-LevelStatistics}). This is indeed expected deep in the MBL phase, where the localized nature of eigenstates leads to a vanishing level repulsion between the system in the limit of large system sizes \cite{PalHuse}.

In a different setting, an all-to-all coupled system of ten superconducting qubits, characterized by disordered XY-spin Hamiltonian, was probed \cite{xu2018}. An initial N{\'e}el state, in close connection to the density wave prepared in cold atom experiments, was used to monitor the dynamical evolution of the system for different disorder strengths. The system exhibited evolution into a steady state, with finite staggered magnetization (imbalance) after a few interaction times, indicating a localized phase of spins. Using quantum state tomography, it was also possible to map out the entanglement entropy of a five-qubit subsystem, which exhibited signs of logarithmic entanglement entropy growth (see Section \ref{Sec:MBL-dyn}).

\subsection{MBL in real materials}
In real materials the coupling between phonons and electrons renders the observation of MBL very difficult. However, promising signatures of localization were reported in thin films of a:InO, which at high magnetic fields undergo a superconductor to insulator transition. For temperatures below 100\,mK and magnetic fields between $0.5<B<2$\,T the system exhibited a 
dramatic drop in conductivity as the temperature was lowered. The conductivity data was found to be incompatible with a simple exponential activation but could, instead, be fitted with a functional form that indicated a critical temperature below which the conductivity vanishes \cite{Shahar}. Additional results seem to indicate that the origin of this vanishing conductivity is intimately connected to a decoupling between electron and phonon temperatures in the system. Both these observations are compatible with theoretical descriptions of an MBL scenario, however, the microscopic origins of the indirectly measured decoupling of electron and phonon temperatures remains unclear so far. While more research seems needed to clarify these questions, the striking results are the most promising ones so far for the existence of MBL in real materials.

Separate experiments using NMR probes of nuclear spin chains showed evidence for growing correlations in an interacting localized system as a function of evolution time \cite{Cappellaro2018}. The results thereby provided additional support for the entangling evolution in an MBL systems (see Section \ref{Sec:MBL-dyn}) in strong contrast to the absence of such an evolution in a non-interacting Anderson insulator. 

Yet in another direction, ~\cite{Aeppli} studied a disordered magnet ${\rm LiHo_{\alpha}Y_{1-\alpha}F_4}$ using pump-probe techniques. They observed that low-energy excitations, with energies much smaller than the microscopic spin-spin interaction scale, were very long-lived, suggesting a dramatic slowdown of thermalization of those excitations. While the precise nature of these (most likely collective) excitations and the origin of their slow decay remain to be understood, it is evident that pump-probe experiments with disordered magnets provide a promising setting for exploring the breakdown of thermalization and MBL. 

\begin{figure}[b]
\begin{center}
\includegraphics[width=0.95\columnwidth]{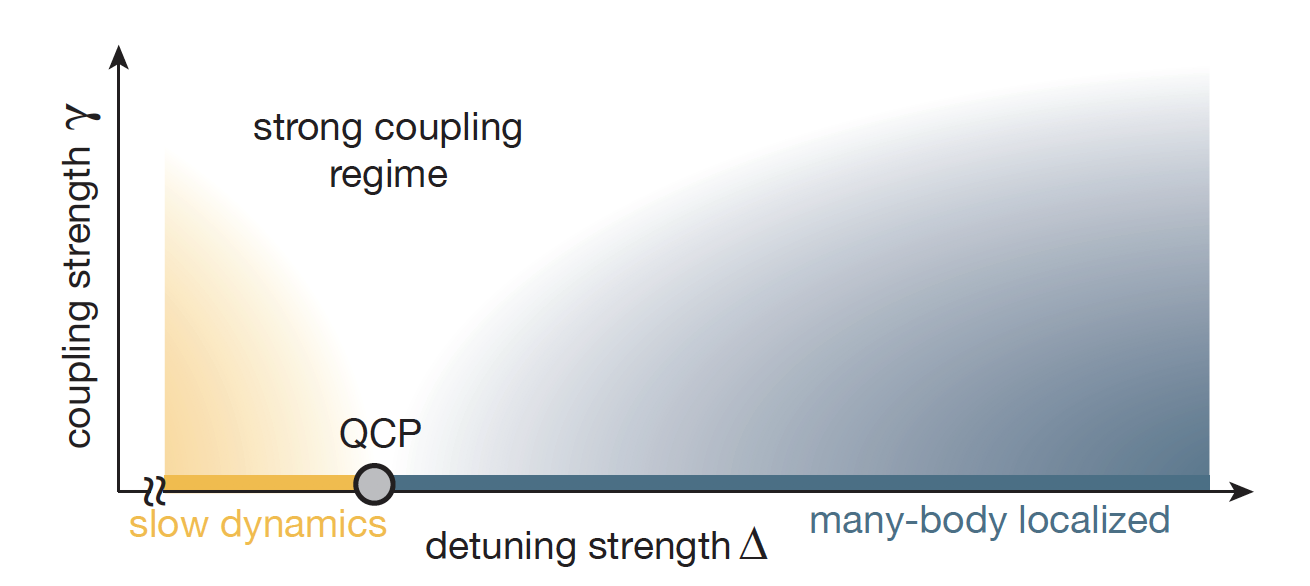}\\
\caption{ \label{Fig:FiniteCoupling}
 Schematic phase diagram of an open MBL system: Coupling an MBL system to a bath destroys the signatures of the system on a timescale that depends on the coupling strength. Signatures might e.g.\ be a persistent imbalance in the MBL phase or a power-law decay of the imbalance in the thermalizing phase. The yellow and blue shaded regions indicate the regimes where those quantities are still accessible. In the white regime of strong couplings, the bath dominates the dynamics of the system. The ideal phases and an actual transition exist only at $\gamma = 0$ (from \cite{lueschen2018}).
  }
\end{center}
\end{figure}

\subsection{Residual coupling to the outside world \label{Sec:open}}

In an ideal MBL scenario, the system is completely isolated from the outside world preventing thermalization with such an infinite size bath. All experiments are, however, to a varying degree, coupled to the external world. How does this affect the observability of different ergodic and localized phases in the system? Let us imagine that the system exhibits a finite coupling strength $\gamma$ to the outside world. Even if $\gamma$ is small, we expect the system to eventually thermalize with the environment for very long time evolutions, thereby destroying any localized phases. For intermediate timescales, however, long compared to any relaxation times in the system and short compared to the coupling to the outside world, we can nevertheless expect the system to exhibit genuine MBL properties. In general, it is therefore suggestive that the situation is rather similar to the case of a finite-temperature quantum phase transition, with the temperature being replaced by the coupling rate $\gamma$ (see Fig.~\ref{Fig:FiniteCoupling}).

In contrast to glassy systems, we expect couplings to thermal reservoirs to have a much stronger  influence on the MBL phase, which thereby also may act as an experimental signature that can distinguish the two from each other. While a classical glass can remain in a glassy phase even when coupled to a reservoir, the MBL phase is expected to be destroyed (either becoming a thermalizing phase, or a glassy phase, in case of a frustrated system at low energy density). Measuring the susceptibility of any MBL phase to external couplings might therefore be a useful probe to distinguish between the two cases \cite{lueschen2017}. In a different direction, \textcite{Lenaric} considered a setup where the system is coupled to phonons and simultaneously subject to a white noise drive, showing that measuring local temperature fluctuations in such a setting provides a way to distinguish MBL from a thermal, and potentially a glassy phase

\section{Outlook \label{Sec:outlook}}

Finally, we close this Colloquium with a discussion of experimental and  theoretical  challenges related to MBL, and to ergodicity breaking in a broader context. 

While there has been significant progress in our ability to describe many aspects of the MBL phase theoretically, many important questions remain open. The transition region remains especially challenging both for theory and experiment. For experiment it is a challenging regime mainly due to the very long time, and possibly also long length scales, on which the critical behavior develops. To capture these long times the experiments must become even better isolated from the environment. It would be also interesting to design experiments to measure the evolution of the entanglement entropy \cite{islam2015}, or to monitor the evolution of fluctuations in the system as an alternative measure of many-body entanglement \cite{Serbyn14}.

\begin{figure}[t]
\begin{center}
\includegraphics[width=0.85\columnwidth]{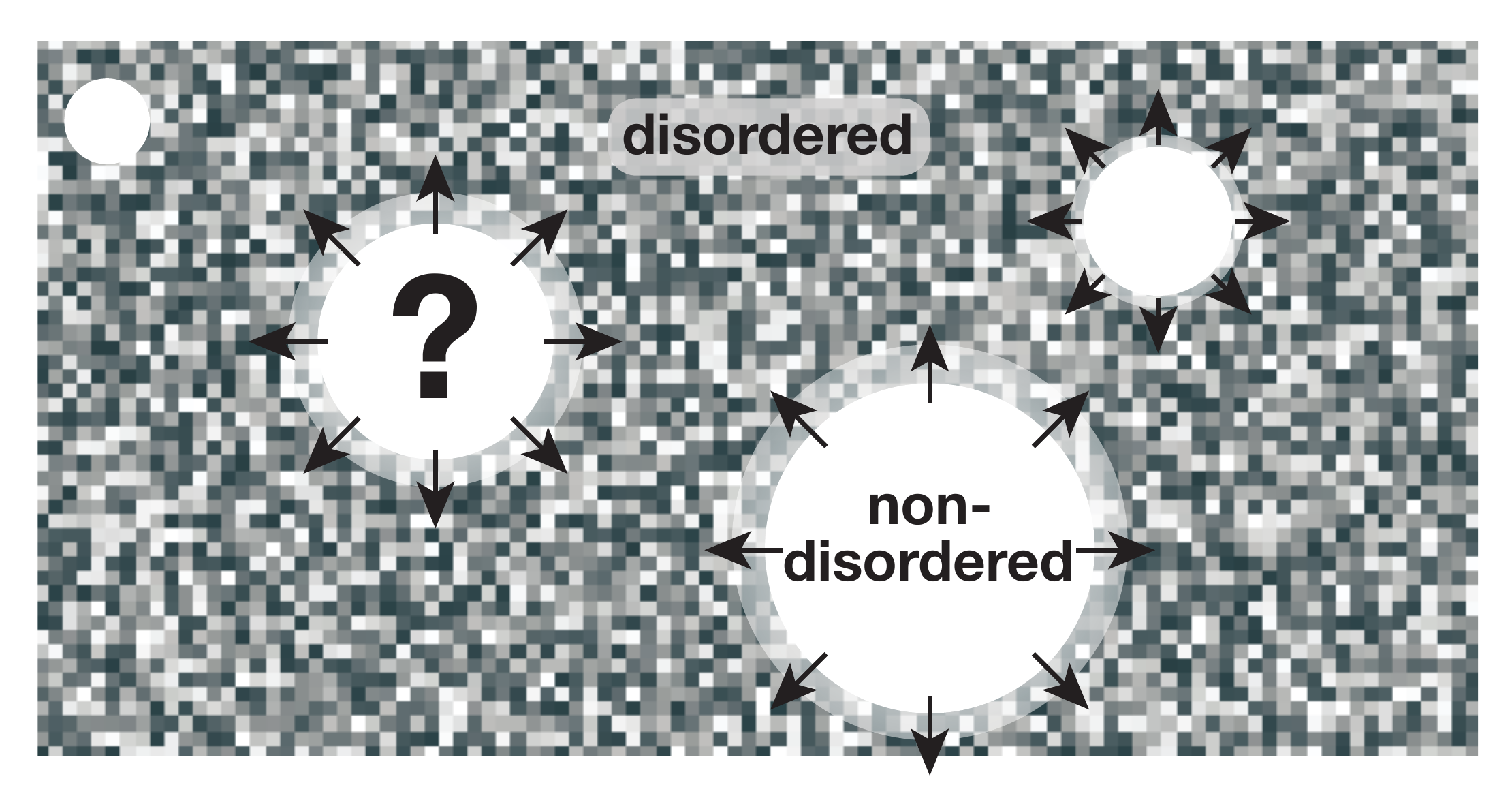}\\
\caption{Engineered disorder pattern. Different techniques allow one to engineer disorder potentials in highly controllable ways. It is, for example, feasible to include non-disordered regions in the disorder pattern, where the system can locally thermalize. The stability of the overall MBL upon different densities and sizes of such thermalizing inclusions could thereby be investigated. This would enable to probe the stability of MBL in fundamentally new ways. \label{Fig:StructuredDisorder}
  }
\end{center}
\end{figure}

Another question concerns the existence of MBL in higher dimensions $d>1$. It has been argued that in this case small thermal inclusions can trigger avalanches in the system that destroy the localized phase~\cite{Roeck,luitz2017,ponte2017}.  The essence of the argument is that a small inclusion thermalizes its immediate neighborhood, thereby becoming a larger bath. If the number of spatial dimensions is larger than one, the incipient bath continues to grow and ultimately thermalizes the entire system. These arguments rest on the applicability of random matrix theory at every stage of the avalanche, which is a subject of debate. Moreover, it is not clear how to reconcile this argument with the approach of \textcite{Basko06} which predicts the existence of an MBL phase in arbitrary spatial dimension. Experiments can help to shed light on this fundamental question by  using structured disorder patterns, where the disorder is interrupted by small non-disordered, thermalizing regions whose density and size can be tuned at will, see Fig.~\ref{Fig:StructuredDisorder}.

Along similar lines, one can ask how stable MBL is when coupled to an ergodic system of approximately the same size \cite{nandkishore2015many,li2015,hyatt2017}. Does the disordered system localize the ergodic one (which plays the role of a bath) or does the combined system become fully ergodic? Could both phases even coexist when the system and the bath are strongly coupled to each other? A natural setting in which localized and delocalized states coexist is a non-interacting system with a single-particle mobility edge (SPME)~\cite{li2015,xiaoli2017}. Such a system has recently been realized in experiment~\cite{lueschen2017spme}. Very weak interactions are expected to immediately drive such a system into an ergodic phase. In contrast, numerical results at relatively strong interactions show indications of localization in this setting~\cite{li2015}. While it is conceivable that strong interactions may localize the initially delocalized single-particle states above the SPME, more detailed theoretical studies are needed. Experiments can again help to shed light on the competition between localization and thermalization in regimes which are inaccessible for numerics. 

 In a broader context,  an exciting challenge for both theory and experiment is to establish whether MBL provides the only robust mechanism of ergodicity-breaking in quantum systems.  MBL systems with their description in terms of LIOMs and a simple area-law entanglement structure of eigenstates provide a useful starting point for addressing  this question. In particular, is it natural to ask if there could exist a system which exhibits only a partial (rather than a complete as in MBL phase) set of quasi-local integrals of motion with a number of LIOMs scaling as some fraction of the total number of physical spins. A possible example of such partial integrability could be provided by systems with many-body mobility edges. It is highly desirable to extend the theory based on LIOMs to such systems~\cite{Scott17}. 
 
Turning to the area-law entanglement structure of MBL eigenstates, one can ask if the  breakdown of thermalization can occur for eigenstates which violate area-law entanglement scaling? As discussed in Section~\ref{sec:SPT}, disordered systems with continuous non-Abelian symmetries, such as SU(2) symmetry, must have logarithmic scaling of entanglement and thus may exhibit new kinds of ergodicity-breaking phases with only partial integrability. 
  
To make progress in the challenges outlined above, new theoretical and numerical methods are needed. In particular, tensor-network based methods for studying the dynamics and eigenstates have to be significantly improved to allow studies of large quantum systems. On the analytical side, extending real-space RG methods to include multi-spin processes that are typically neglected appears to be a promising direction. 

To conclude, theoretical and experimental advances in many-body localization have revealed a new universality class of quantum dynamics, opening up a new frontier in non-equilibrium physics. As we discussed above, much of the theory progress was inspired by applying concepts from quantum information theory to non-equilibrium phenomena. On the experimental side, studies of MBL in quantum systems were made possible by the remarkable technological progress in realizing synthetic quantum systems. Looking forward, a close collaboration between experiment and different branches of theory will almost certainly lead to the discovery of new non-equilibrium states with unexpected and potentially useful properties.

\section*{Acknowledgments}

We are grateful to all our collaborators on the topic of this review, including Guifre Vidal, Eugene Demler, Misha Lukin, Wen Wei Ho, Soonwon Choi, Francois Huveneers, Wojciech De Roeck, Isaac Kim, Ulrich Schneider, Christian Gross, Ronen Vosk, Michael Knap, Ivan Protopopov, Anushya Chandran, Vedika Khemani, Sarang Gopalakrishnan, Norman Yao, Chris Laumann, Joel Moore, and especially Zlatko Papi\'c and David Huse.  We acknowledge many enlightening discussions with Boris Altshuler, Jens Bardarson, Leticia Cugliandolo, Thierry Giamarchi, Achilleas Lazarides, Sasha Mirlin, Roderich Moessner, Rahul Nandkishore, Arijet Pal, Sid Parameswaran, Anatoli Polkovnikov, Frank Pollmann, Andrew Potter, Sankar Das Sarma, Antonello Scardicchio, Shivaji Sondhi, Romain Vasseur, and Ashvin Vishwanath.

\end{document}